\PassOptionsToPackage{square,comma,numbers,sort&compress,super}{natbib}
\documentclass[12pt]{article}
\usepackage{amsthm, amscd, amsfonts, amssymb, graphicx,tikz, color, environ}
\usepackage{amsmath}
\usepackage{rotating} 
\usepackage{hyperref}
\usepackage{caption} 
\usepackage{float}
\usepackage{bm}
\usepackage{xcolor}
\usepackage{comment}
\usepackage{cleveref}
\usepackage{xcolor}
\usepackage{ragged2e}
\usepackage[utf8]{inputenc}
\usepackage[export]{adjustbox}
\usepackage{graphicx}

\usepackage{enumitem}
\newcommand\numberthis{\addtocounter{equation}{1}\tag{\theequation}}
\usepackage[sort&compress,numbers,comma,square]{natbib} 
\hypersetup{
   colorlinks, 
    linkcolor=blue, 
  citecolor=blue
}

\begin{document}
\begin{center}{\Large Likelihood-based Missing Data Analysis in Crossover Trials}\end{center}
\begin{center}{{ Savita Pareek\textsuperscript{\color{blue}1}, Kalyan Das\textsuperscript{\color{blue}1}, and Siuli Mukhopadhyay\textsuperscript{\color{blue}1\color{black},} {\footnote[2]{Corresponding author. Email: siuli@math.iitb.ac.in}}}\\
\vspace{.15em}
{\textsuperscript{1}\it Department of Mathematics, Indian Institute of Technology Bombay,}\\ {\it Mumbai 400 076, India}}\end{center}
\date{}
\hrule
\vspace{.3em}
\section*{Abstract}
A multivariate mixed-effects model seems to be the most appropriate for gene expression data collected in a crossover trial. It is, however, difficult to obtain reliable results using standard statistical inference when some responses are missing. Particularly for crossover studies, missingness is a serious concern as the trial requires a small number of participants. A Monte Carlo EM (MCEM)-based technique was adopted to deal with this situation. In addition to estimation, MCEM likelihood ratio tests (LRTs) are developed to test fixed effects in crossover models with missing data. Intensive simulation studies were conducted prior to analyzing gene expression data.

\smallskip
\noindent \textbf{Keywords:} Crossover trials; Monte Carlo EM algorithm; MCEM LRTs;
\section{Introduction}
In many clinical studies, we may encounter crossover trials for two or more response variates with subject dropouts and incomplete data. For example, one may consider the measurement of both systolic (SBP) and diastolic (DBP) blood pressure in each period (\citet{Grender1993}) or the recording of blood sugar levels at multiple time points in each period (\citet{Putt1999}) or microarray gene expression profiles of subjects measured in each period (\citet{Leaker2016}) with incomplete data. To deal with such multivariate crossover trials, we propose a mixed-effect model to account for the multiple responses measured in each period from each subject. To address the incomplete data problem in a multivariate crossover model framework, we use a Monte Carlo EM (MCEM) based estimation method.

%
%
In a crossover trial, usually a smaller number of participants as compared to cross-sectional studies are recruited as the same subjects are switched between all treatment groups. \textcolor{black}{Thus, using ad-hoc missing data analysis methods such as list-wise deletion of subjects with missing responses or imputation-based methods to deal with the subject dropout may lead to highly underpowered studies and biased estimates (\citet{pigott2001}, \citet{Schafer2002}, \citet{briggs2003}).} Similarly, ignoring the correlated nature of the multiple responses recorded in each period and fitting parallel univariate crossover models to each response may lead us to inaccurate conclusions about the intra and inter-subject variabilities. 

For a detailed review of univariate crossover designs, refer to the books by (\citet{senn1993};  \citet{Jones}). In comparison, crossover trials with multiple responses measured in each period have  been addressed by fewer researchers, namely (\citet{Grender1993}; \citet{chin1996}; \citet{Putt1999}; \citet{Tudor2000}). To tackle the presence of missing/incomplete  data specifically for the $AB/BA$ crossover design under the MAR or NMAR missing mechanism, one may refer to (\citet{Patel1985};  \citet{Ho2012}; \citet{Matthews2013}). For more general univariate crossover designs with incomplete data  references are namely, (\citet{Richardson1992}; \citet{Chow1997}; \citet{Basu2010}; \citet{Rosenkranz2015}). However, none of the works cited deal with missingness in a multivariate crossover design setup.

In this article, we model and analyze multivariate crossover studies or studies with multiple response variates in the presence of missing at random outcomes. This work is motivated by a $3\times 3$ crossover trial studying the effect of two single doses of oral prednisone (10 and 25 mg) with a placebo on biomarkers of mucosal inflammation and transcriptomics (\citet{Leaker2016}). Ten gene expression changes of subjects were measured in each trial period, giving rise to multiple responses in a crossover framework. The crossover study involved 17 subjects assigned to three treatment sequences. Several outcomes were missing, particularly in the third period. Our task here is to fit an appropriate statistical model to the crossover trial with micro-array expression data taking into account the multivariate structure of the data and the missing responses. Mixed models based on Monte Carlo EM are used to fit and estimate treatment and gene effects and intra- and inter-subject variability in the presence of MAR outcomes. The mixed model, assumed, include the fixed effects of treatment, period, and genes and their interactions, while the random effects are subject-specific effects. We use maximum likelihood estimation coupled with the MCEM algorithm to determine the estimated model parameters and variance components. The asymptotic covariance matrix is obtained by multiple imputations (\citet{Goetghebeur2000}). MCEM estimators are used to formulating likelihood ratio tests (LRTs). Detailed simulations to study the properties of the estimators and the power of the proposed LRTs have been provided. An example illustrating the statistical model and its estimation is based on the gene expression levels measured in a $3\times 3$ crossover study.

To the best of our knowledge, this is the first work involving multivariate crossover studies with missing responses. Though originally developed for agricultural sciences, crossover trials are nowadays frequently used in clinical trials and biological studies to evaluate treatment effects. These trials are most useful for comparing treatments for chronic or long-term diseases, such as asthma, rheumatism, hypertension, etc. Note that one useful advantage of crossover studies is that they require a smaller number of participants. However, due to the smaller number of recruits, reliable statistical inference is harder to make when some responses are randomly missing.

\section{Case Study: Multivariate Crossover Trial of Oral Prednisone}\label{case study}
\begin{figure}[!htp]
\centering
  \includegraphics[width=13cm]{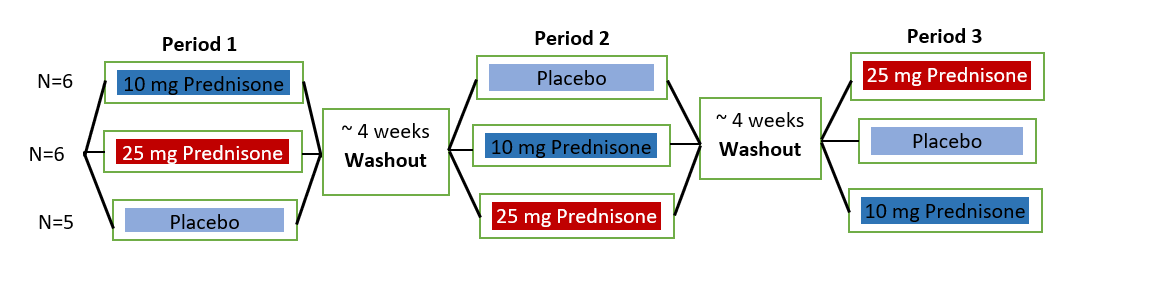}
  \caption{{ Study design: a three-way crossover trial to examine the effects of a single oral
dose of prednisone (10 or 25 mg) versus placebo given before nasal allergen challenge (NAC). Picture is reproduced from \citet{Leaker2016}}}
  \label{fig:study design}
 \end{figure}

We use a gene expression dataset from (\citet{Leaker2016}) as a case study. The dataset is publicly available from the NCBI Gene Expression Omnibus (\citet{Clough2016}) and can be accessed using the hyperlink, \href{https://www.ncbi.nlm.nih.gov/geo/query/acc.cgi?acc=GSE67200}{nasal mRNA data}. 
The gene expression study results from a  randomized, double-blind, placebo-controlled, three-period, crossover trial. It is to assess the effects of two single doses of oral prednisone (10 mg, 25 mg) on inflammatory mediators measured in nasal exudates after a nasal allergen challenge in susceptible individuals with allergic rhinitis.
All subjects have a history of seasonal asthma rhinitis to grass pollen and a positive result from the intraepidermal skin prick test to grass pollen extract. The study involves seventeen subjects assigned to three treatment sequences/groups. There are also some missing observations (particularly in the third period). The main objective of this study is to determine the effect of treatments and genes on allergic reactions to grass pollen. The outcomes measured are a fold change of mRNA expression levels (pg ml$^{-1}$), \textit{i.e.,} changes in gene expression values for ten genes recorded in the nasal allergen challenge. 
\Cref{fig:study design} and \Cref{tab:mis responses} describe the study design.
 \begin{table}[!htp]
 \scriptsize
 \begin{center}
 \caption{\textcolor{black}{Gene data: the number of missing subjects, periods, and genes in each sequence.}}
 \label{tab:mis responses}
\begin{tabular}{lllll} \hline
& Sequence 1          & Sequence 2           & Sequence 3            & Overall \\ \hline
Total subjects          & 6 (Subjects 1 to 6) & 6 (Subjects 7 to 12) & 5 (Subjects 13 to 17) & 17      \\
No. of missing subjects & 2 (Subjects 2, 5)   & 2 (Subjects 8, 10)   & 1 (Subject 14)        & 5       \\
No. of missing period   & 1 (Period 3)        & 1 (Period 3)         & 1 (Period 3)          & 3       \\
No. of missing genes    & 10 (Genes 1 to 10)  & 10 (Genes 1 to 10)   & 10 (Genes 1 to 10)    & 30      \\
Percent of missingness  & 11.1\%                & 11.1\%                 & 6.7\%                   & 9.8\%  \\ \hline
\end{tabular}
 \end{center}
\end{table}

For instance, Sequence 1 indicates that 10 mg of prednisone will be administered to subjects (1 to 6) in period 1. Following a washout period of four weeks, a placebo will be administered to subjects at period 2, 4 weeks later a prednisone dose of 25 mg will be administered to each subject at period 3. 
For various sequences, there are missing responses in period 3. 
The overall missing response rate is {$9.8\%$}. The data set is described in more detail in \Cref{tab:mis responses}.

\begin{figure}[!htp]
\centering
  \includegraphics[width=10cm,height=9cm]{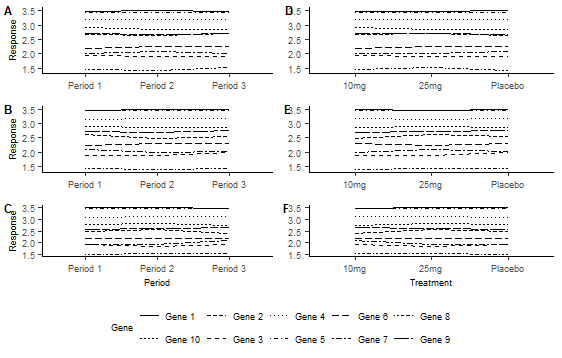}
  \caption{\textcolor{black}{Period by gene and treatment by gene interaction plots: Plots on the left (A, B, C) are period by gene interaction plots for sequences 1, 2, and 3, respectively. Plots on the right (D, E, F) are treatment by gene interaction plots for sequences 1, 2, and 3, respectively.}}
  \label{fig:period trt by gene}
 \end{figure}

\begin{figure}[!htp]
\centering
  \includegraphics[width=10cm,height=6cm]{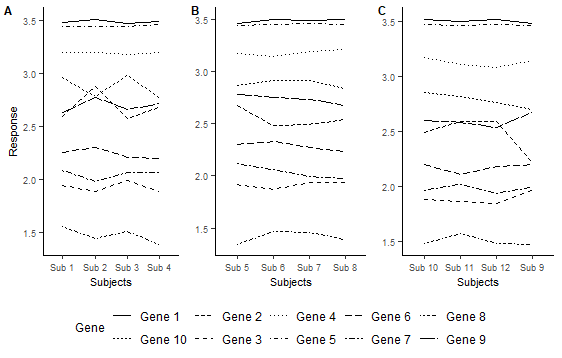}
  \caption{ \textcolor{black}{Subject by gene interaction plot: Plots A, B, and C represent the subject by gene interaction plots for sequences 1, 2, and 3, respectively.}}
  \label{fig:subject by gene}
 \end{figure}

Before model fitting, we ran some exploratory analysis on the gene data based only on complete cases. A complete case in the crossover setup means that a subject who missed an observation in any period has been totally removed. From \Cref{tab:mis responses}, we see two missing subjects in sequences 1, and 2 and one missing subject in sequence 3. Consequently, the exploratory analysis is based on four subjects per sequence. In \Cref{fig:period trt by gene} and \Cref{fig:subject by gene}, gene by treatment, gene by period, and subject by gene interactions were seem to be almost absent. 

\textcolor{black}{We also tried to assess whether the outcomes are missing completely at random (MCAR) using Little's MCAR test (\citet{littletest}). 
The p-value is close to zero, which indicates that the responses are not MCAR, but could be MAR or not missing at random (NMAR). We assume throughout this manuscript that missingness in the responses is due to MAR. In \Cref{sec mle missing}, we explain why we consider MAR to be a mechanism for dealing with missing data.
Moreover, the response variable can have missing values in a crossover study in several ways. An individual subject can have all the responses unavailable from a particular period or have missing responses in different periods resulting in a monotone or non-monotone pattern of missingness.}

 \textcolor{black}{ Thus, in the following section, we present a specific model that fits a sample of the microarray expression data outlined above, assuming that some responses are missing at random (MAR) from this sample.} 

\section{Model and Notation}\label{modelnotations}
 Suppose $y_{ijkl}$ denotes measurement of the $l^{\text{th}}$ gene expression from the $k^{\text{th}}$ subject in the $j^{\text{th}}$ period assigned to the $i^{\text{th}}$ treatment sequence, where $i = 1 (1) s$;  $j = 1 (1) p$; $k = 1 (1) n_i$ and $l = 1 (1) m$. So there are in total $n$ subjects, where $n=\sum_{i=1}^{s}n_i$. Assuming a normal mixed effects model, we may write,
\newcommand\iid{i.i.d.}
\newcommand\pN{{N}}
\begin{equation}\label{eq:1}
y_{ijkl}= \mu + \pi_{j}+\tau_{d[i,j]}+g_{l}+s_{ik}+e_{ijkl}
\end{equation} 
where 
$\mu$ is  an intercept, 
$\pi_{j}$ is the fixed effect of the $j$th period,
$\tau_{d[i,j]},$ $(d[i,j]= 1 (1) t)$ is the  fixed effect associated with the treatment applied in period $j$ of the sequence $i$,  
$g_{l}$ is the fixed effect of the $l^{\text{th}}$ gene,
$s_{ik}$ is a random effect associated with the $k^{\text{th}}$ subject in sequence $i$, and $e_{ijkl}$ is  a random error. We assume that,  $e_{ijkl} \overset{ind}{\sim} \pN(0, \sigma_{e}^2)\;\; \forall\, i,j,k,l$,  and $s_{ik} \overset{ind}{\sim} \pN(0, \sigma_{s}^2)\;\; \forall\, i,k$. Also, $s_{ik}, e_{ijkl}$ are independent. Note, we have not considered an interaction effect of gene with period and treatment in the above model, in accordance with the exploratory analysis results (see \Cref{fig:period trt by gene}) for the case study data. 
Based on the above assumptions, we have $\text{var}(y_{ijkl})=\sigma_{e}^2+\sigma_{s}^2$, and $\text{cov}(y_{ijkl},y_{ij^{'}kl^{'}} )=\sigma_{s}^2$ for all $ j \neq j^{'}$, $l \neq l^{'}$. Thus, variance-covariance matrix corresponding to the $k^{\text{th}}$ subject administered the $i^{\text{th}}$ treatment sequence is given by,
$$var(\bm{y}_{ik}) =(\bm{ \Sigma}_{ik})_{pm \times pm} = \begin{bmatrix}
 \sigma_{e}^2 \bm{R} & \sigma_{s}^2\bm{ 1}_{m}\bm{1}_{m}^{T} &.&.&.& \sigma_{s}^2 \bm{1}_{m}\bm{1}_{m}^{T} \\
\sigma_{s}^2 \bm{1}_{m}\bm{1}_{m}^{T} &  \sigma_{e}^2\bm{ R} &.&.&.& \sigma_{s}^2 \bm{1}_{m}\bm{1}_{m}^{T} \\
\vdots\\
\sigma_{s}^2 \bm{1}_{m}\bm{1}_{m}^{T}  & \sigma_{s}^2 \bm{1}_{m}\bm{1}_{m}^{T}  &.&.&.&  \sigma_{e}^2 \bm{ R}
\end{bmatrix}$$ 
where, $\bm{y}_{ik}=(y_{i1k1},\ldots, y_{ipkm})_{pm \times 1}^{T}$, $\bm{R}=\bm{I}_{m}+\frac{\sigma_{s}^2}{\sigma_{e}^2}\bm{1}_{m}\bm{1}_{m}^{T}$ and $\bm{1}_{m}^{T}=(1,1,\ldots,1)$. 
 
Using matrix notations we rewrite \cref{eq:1} as 
\begin{equation}\label{eq:2}
\bm{y}_{i}=\bm{ X}_{i}\bm{\beta}+\bm{Z}_{i}\bm{b}_{i}+\bm{e}_{i},~ i=1 (1) s,
\end{equation}
where, 
\begin{itemize}
\item[(i)] $\bm{y}_{i}=(\bm{y}_{i1}^{T},\bm{y}_{i2}^{T},\ldots,\bm{y}_{ik}^{T},\dots,\bm{y}_{in_{i}}^{T})^T$ is a response vector of length $pmn_{i}$,
\item[(ii)] $\bm{X}_{i}=(\bm{X}_{i1}^{T},\bm{X}_{i2}^{T},\ldots,\bm{X}_{ik}^{T},\ldots,\bm{X}_{in_{i}}^{T})^{T}_{pmn_{i}\times (p+t+m-2)}$ is the design matrix corresponding to the fixed effects, and  $$\bm{X}_{ik}=(\bm{1}_{p}\otimes \bm{1}_{m},((\bm{I}_{p-1}\otimes \bm{1}_{m})^{T},\bm{0}_{m \times (p-1)}^{T})^{T}_{pm \times (p-1)},\bm{T}_{pm \times (t-1)},\bm{1}_{p}\otimes(\bm{I}_{m-1},\bm{0}_{m-1})^{T}).$$  The ${pm \times (t-1)}$ matrix of treatment effects is represented as $\bm{T}=(T_{1}^{T}, \ldots,T_{p}^{T})^{T}$, where $ T_{u}=(\bm{1}_{m} \otimes (a_{1},\ldots, a_{t-1}))_{m \times (t-1)}$ $\text{for}\; u=1 (1) p$, and $a_l$ is an indicator variable which takes value $1$ if the $l^{\text{th}}$ treatment is assigned to the $u^{\text{th}}$ period and $0$ otherwise. 
\item[(iii)] The parameters of interest are,\\
 $\bm{\beta}=(\mu , \pi_{1} ,\dots \pi_{p-1},  \tau_{1},\ldots,\tau_{t-1}, g_{1},\ldots,g_{m-1})$ and $\sigma^2 =(\sigma_e^2,\sigma_s^2).$
 \item[(iv)] $\bm{Z}_{i}=\bm{I}_{n_{i}} \otimes \bm{1}_{pm}$ is the design matrix corresponding to the random effects.
 \item[(v)] Moreover,  $\bm{b}_{i}=(s_{i1},s_{i2}, \ldots, s_{in_{i}})^{T}$ is the ${n_{i}\times 1}$ vector of random effects, and  $\bm{e}_{i}=(\bm{e}_{i1}^{T},\bm{e}_{i2}^{T},\ldots,\bm{e}_{iq}^{T},\ldots,\bm{e}_{in_{i}}^{T})^{T}$ is the ${pmn_{i} \times 1}$ random error vector, where $\bm{e}_{iq}=(e_{i1q1}, \ldots ,e_{ipqm})^{T}$, it is assumed \begin{equation*}
\bm{e}_{i} \sim \pN(\bm{0},\sigma_{e}^2 \bm{I}_{pmn_{i}}); \;\; \bm{b}_{i}  \sim \pN_{n_{i}}(\bm{0},\bm{D}_{i}), \; \bm{D}_{i}=\sigma_{s}^2\bm{I}_{n_{i}}.
\end{equation*}

 \end{itemize}

Under the above assumptions, the conditional distribution of $ \bm{y}_{i}$ given the random effects $\bm{b}_{i}$  is of the form \begin{equation*}\label{eq:3}
 \bm{y}_{i}|\bm{\beta},\sigma_{e}^2,\bm{b}_{i}\sim \pN_{pmn_{i}}(\bm{X}_{i}\bm{\beta}+\bm{Z}_{i}\bm{b}_{i},\sigma_{e}^2\bm{ I}_{pmn_{i}}),
 \end{equation*}\noindent while the marginal distribution of $\bm{y}_{i}$ is \begin{equation*}\label{eq:4}
 \bm{y}_{i}|\bm{\beta},\sigma_{e}^2,\bm{D}_{i} \sim \pN_{pmn_{i}}(\bm{X}_{i}\bm{\beta}, \bm{\Sigma}_{i}), \; \bm{\Sigma}_{i}=\bm{Z}_{i}\bm{D}_{i}\bm{Z}_{i}^{T}+\sigma_{e}^{2}\bm{I}_{pmn_{i}}.
 \end{equation*}

\section{Estimation with Missing at Random Responses}\label{sec mle missing}

Here, we present a maximum likelihood estimation approach for crossover design with multiple responses and missing observations. In the earlier section, we saw that the crossover model is a  special situation of the linear mixed model. To accommodate the multiple responses observed in each period in a crossover setup, the structure of the covariance matrices of the linear mixed model became more complex. Moreover, we noted in the motivating example that some responses have missing values. Sometimes an individual subject may have all the responses unavailable from a particular period, or responses may be missing from different periods. To account for the missing responses we assume the MAR mechanism, which may be monotone or non-monotone in nature, and use the MCEM algorithm to fit the data. 

\textcolor{black}{Before beginning the estimation process, it is essential to understand the existing types of missing data. 
There are three types of missing data mechanisms defined in \citet{rubin}: missing completely at random (MCAR), missing at random (MAR), and not missing at random (NMAR). A missing data mechanism is known as MCAR when missing data are not correlated with either the missing responses or the observed values. MAR data are those in which the probability of an observation missing depends only on data that has been observed. In the case of NMAR, the probability of missing depends on the observed data and the missing observations, implying that the probability of missing may vary for unidentified reasons. Further, missing observations may occur monotonically or non-monotonically. The term monotone missingness in the context of the motivating example implies that if a particular gene expression level is not observed in a given time period, all subsequent observations are also absent. On the other hand, non-monotone missingness implies that a particular gene expression level may be missing for a specific time period, but some subsequent observations will still be observed.}
\textcolor{black}{In likelihood-based methods, MCAR and MAR missing data mechanisms are considered \textit{ignorable}. It implies that a parametric form of missing data mechanism does not need to be explicitly addressed in the analysis. Alternatively, the NMAR missing data mechanism is considered \textit{non-ignorable} in that the parametric model for the missing data mechanisms must be incorporated into the complete data log-likelihood (\citet{Stubbendick2003}, \cite{Stubbendick2006} and \citet{Ibrahim2009}). In the NMAR case, two ways have been identified for factoring the complete data log-likelihood: selection and pattern-mixture models (\citet{Little1995}).}

\textcolor{black}{When no information is available about the causes of missingness or nonresponse among subjects, researchers usually assume MAR as a starting point for their analysis. In cancer and HIV trials, unobserved biomarkers often cause patients to become unresponsive as the treatment is not effective, causing them to drop out, implying an NMAR missing mechanism. However, since the case study discussed here is simply an allergen challenge, we do not have access to the kind of information collected in cancer trials, so MAR is used as a starting point.}

We use the MCEM algorithm of (\citet{Wei1990}, \citet{casella2001}) for parameter estimation. \textcolor{black}{To make the manuscript self-sufficient, we have included a short description of the MCEM algorithm and some aspects of its convergence in the Appendix (\Cref{secmcem})}. \textcolor{black}{ In general, the method described next based on the MCEM algorithm can be applied to both monotonic and non-monotonic missing data sets.} 

\textcolor{black}{
Suppose, $\bm{y}_{i}=(\bm{y}_{mis,i}^{T},\bm{y}_{obs,i}^{T})^{T}$ where $\bm{y}_{mis,i}$ is the $m_{i} \times 1$ vector  of missing responses, $\bm{y}_{obs,i}$ is the $l_{i} \times 1$ vector  of observed responses. The sum of $m_{i}$ and $l_{i}$ is equal to $pmn_{i}$. The missing values are assumed to be in the responses or outcomes, whereas the covariates were all observed.}

\textcolor{black}{ To execute the EM algorithm, we followed the steps listed in (\citet{Ibrahim2009}).}
 \textcolor{black}{
In the E-step for the $i^{th}$ sequence at the $(t+1)^{th}$ iteration we compute,
\begin{align*}
Q_{i}(\bm{\gamma}|\bm{\gamma}^{(t)})&=E\left[l(\bm{\gamma};\bm{y}_{i},\bm{b}_{i}|\bm{y}_{obs,i} \bm{\gamma}^{(t)})\right]\\
&=\int \int \log \left[f(\bm{y}_{i}|\bm{\beta},\sigma_{e}^2,\bm{b}_{i},) \right]f(\bm{y}_{mis,i},\bm{b}_{i}|\bm{y}_{obs,i},\bm{\gamma}^{(t)}) d\bm{b}_{i}  d\bm{y}_{mis,i}\\
&+\int \int \log \left[f(\bm{b}_{i}|\bm{D}_{i}) \right]f(\bm{y}_{mis,i},\bm{b}_{i}|\bm{y}_{obs,i},\bm{\gamma}^{(t)}) d\bm{b}_{i}  d\bm{y}_{mis,i}\\
& \equiv I_{1}+I_{2},\numberthis \label{eq:qfn}
\end{align*}
where $\bm{\gamma}^{(t)}=(\bm{\beta}^{(t)},\sigma_{e}^{2(t)},\sigma_{s}^{2(t)})$. 
The conditional distribution, $f(\bm{y}_{mis,i},\bm{b}_{i}|\bm{y}_{obs,i},\bm{\gamma}^{(t)})$, can be written as
\begin{align*}\label{eqestep}
f(\bm{y}_{mis,i},\bm{b}_{i}|\bm{y}_{obs,i},\bm{\gamma}^{(t)})&=f(\bm{b}_{i}|\bm{y}_{i},\bm{\gamma}^{(t)})f(\bm{y}_{mis,i}|\bm{y}_{obs,i},\bm{\gamma}^{(t)}).
\end{align*}
Using standard calculations we obtain $\bm{b}_{i}|\bm{y}_{i}$,} where
\begin{equation}\label{b giv y}
\bm{b}_{i}|\bm{y}_{i},\bm{\gamma}^{(t)}  \sim N(\bm{b}_{0i}^{(t)},\bm{ \Sigma}_{0i}^{(t)})\text{ and, }
\end{equation} 
\begin{align*}
\bm{b}_{0i}^{(t)}&=\bm{D}_{i}^{(t)}\bm{Z}_{i}^{'}(\bm{Z}_{i}\bm{D}_{i}^{(t)}\bm{Z}_{i}^{'}+\sigma_{e}^2 \bm{I}_{pmn_{i}})^{-1}(\bm{y}_{i}-\bm{X}_{i}\bm{\beta}^{(t)})\\
&=\bm{\Sigma}_{0i}^{(t)}\bm{Z}_{i}^{'}(\bm{y}_{i}-\bm{X}_{i}\bm{\beta}^{(t)})/\sigma_{e}^{2(t)}\\
\bm{\Sigma}_{0i}^{(t)}&=\bm{D}_{i}^{(t)}-\bm{D}_{i}^{(t)}\bm{Z}_{i}^{'}(\bm{Z}_{i}\bm{D}_{i}^{(t)}\bm{Z}_{i}^{'}+\sigma_{e}^2 \bm{I}_{pmn_{i}})^{-1}\bm{Z}_{i}\bm{D}_{i}^{(t)}\\
&=\left[\sigma_{e}^{-2(t)}\bm{Z}_{i}^{'}\bm{Z}_{i}+(\bm{D}_{i}^{(t)})^{-1}\right]^{-1}.
\end{align*}
As a result $I_1$ and $I_2$ reduce to,
\begin{align*}
I_{1}&=\frac{-pmn_{i}}{2} \log 2\pi -\frac{pmn_{i}}{2}\log\sigma_{e}^2- \frac{1}{2 \sigma_{e}^2}Tr(\bm{Z}_{i}^{T}\bm{Z}_{i}\bm{\Sigma}_{0i}^{(t)})\\&-\frac{1}{2 \sigma_{e}^2}\int(\bm{y}_{i}-\bm{X}_{i}\bm{\beta}-\bm{Z}_{i}\bm{b}_{0i}^{(t)})^{T}\\&(\bm{y}_{i}-\bm{X}_{i}\bm{\beta}-\bm{Z}_{i}\bm{b}_{0i}^{(t)})f(\bm{y}_{mis,i}|\bm{y}_{obs,i},\bm{r}_i, \bm{\gamma}^{(t)}) d\bm{y}_{mis,i}.\numberthis \label{eq:first int}
\end{align*} and,
\begin{align*}
I_{2}&=\frac{-(m+1)n_{i}}{2} \log 2\pi -\frac{1}{2} \log(\det \bm{D}_{i})-\frac{1}{2} Tr(\bm{D}_{i}^{-1}\bm{\Sigma}_{0i}^{(t)})\\&-
\frac{1}{2} \int({\bm{b}_{0i}^{(t)}}^T\bm{D}_{i}^{-1}\bm{b
}_{0i}^{(t)})f(\bm{y}_{mis,i}|\bm{y}_{obs,i}, \bm{r}_i,\bm{\gamma}^{(t)}) d\bm{y}_{mis,i}.\numberthis \label{eq:second int}
\end{align*}
\textcolor{black}{
To solve the above integrals with respect to $\bm{y}_{mis,i}$ in \cref{eq:first int} and \cref{eq:second int}, we require $f(\bm{y}_{mis,i}|\bm{y}_{obs,i})$.
A Monte Carlo average of independent samples of $[\bm{y}_{mis,i}|\bm{y}_{obs,i},\bm{\gamma}^{(t)}]$ is calculated to approximate the integrals $I_1$ and $I_2$. For the same, we generate independent samples $v_{i1},v_{i2},\ldots v_{ic_{i}}$ of size $c_i$ from $[\bm{y}_{mis,i}|\bm{y}_{obs,i},\bm{\gamma}^{(t)}]$. \textcolor{black}{Appendix (\Cref{secapp}) gives a detailed description of generating these samples.} Using the data $\bm{y}_{i}^{(k)}=(\bm{v}_{ik}^{T},\bm{y}_{obs,i}^{T})^{T}$ and $\bm{b}_{0i}^{(tk)}=\bm{\Sigma}_{0i}^{(t)}\bm{Z}_{i}^{T}(\bm{y}_{i}^{(k)}-\bm{X}_{i}\bm{\beta}^{(t)})\sigma_{e}^{2(t)}$, \cref{eq:first int}, \cref{eq:second int}, the $E$-step for the $i^{\text{th}}$ sequence (\cref{eq:qfn}) in the $(t+1)^{\text{th}}$ iteration then takes the form
\begin{align*}
Q_{i}(\bm{\gamma}|\bm{\gamma}^{(t)})&=\frac{-pmn_{i}}{2} \log 2\pi -\frac{pmn_{i}}{2}\log\sigma_{e}^2- \frac{1}{2 \sigma_{e}^2}Tr(\bm{Z}_{i}^{T}\bm{Z}_{i}\bm{\Sigma}_{0i}^{(t)})\\&-  \frac{1}{2 c_i \sigma_{e}^2} \displaystyle\sum_{k=1}^{c_{i}}(\bm{y}_{i}^{(k)}-\bm{X}_{i}\bm{\beta}-\bm{Z}_{i}\bm{b}_{0i}^{(tk)})^{T}(\bm{y}_{i}^{(k)}-\bm{X}_{i}\bm{\beta}-\bm{Z}_{i}\bm{b}_{0i}^{(tk)})\\
&-\frac{(m+1)n_{i}}{2} \log 2\pi -\frac{1}{2} \log(\det \bm{D}_{i})-\frac{1}{2} Tr(\bm{D}_{i}^{-1}\bm{\Sigma}_{0i}^{(t)})\\&-
\frac{1}{2c_{i}} \displaystyle \sum_{k=1}^{c_{i}}{\bm{b}_{0i}^{(tk)}}^T \bm{D}_{i}^{-1}\bm{b}_{0i}^{(tk)}.
\end{align*}
}
 The E-step for all the sequences is given by
$$Q(\bm{\gamma}|\bm{\gamma}^{(t)})=\displaystyle \sum_{i=1}^{s}Q_{i}(\bm{\gamma}|\bm{\gamma}^{(t)}).$$

In the M-step, we maximize $Q(\bm{\gamma}|\bm{\gamma}^{(t)})$ to obtain
 \begin{align*}
\bm{\beta}^{(t+1)}&=\bigg(\displaystyle\sum_{i=1}^{s} \bm{X}_{i}^{T}\bm{X}_{i}\bigg)^{-1} \displaystyle\sum_{i=1}^{s}\bm{X}_{i}^{T}\frac{1}{c_{i}}  \displaystyle\sum_{k=1}^{c_{i}}(\bm{y}_{i}^{(k)}-\bm{Z}_{i}\bm{b}_{0i}^{(tk)}).\\
\sigma_{e}^{2(t+1)}&=\frac{1}{pmn}\displaystyle\sum_{i=1}^{s}\bigg[\frac{1}{c_{i}} \displaystyle\sum_{k=1}^{c_{i}}(\bm{y}_{i}^{(k)}-\bm{X}_{i}\bm{\beta}^{(t+1)}-\bm{Z}_{i}\bm{b}_{0i}^{(tk)})^{T}\\&(\bm{y}_{i}^{(k)}-\bm{X}_{i}\bm{\beta}^{(t+1)}-\bm{Z}_{i}\bm{b}_{0i}^{(tk)})
+Tr(\bm{Z}_{i}^{T}\bm{Z}_{i}\bm{\Sigma}_{0i}^{(t)})\bigg].\\
\sigma_{s}^{2(t+1)}&=\frac{1}{n}\displaystyle\sum_{i=1}^{s}\left[ Tr(\bm{\Sigma}_{0i}^{(t)})+
\frac{1}{c_{i}} \displaystyle \sum_{k=1}^{c_{i}}{\bm{b}_{0i}^{(tk)}}^T \bm{I}_{n_{i}}\bm{b}_{0i}^{(tk)}\right].
\end{align*}
The $E$ and $M$ steps are repeated until convergence is reached. 

The asymptotic covariance matrix was determined using the multiple imputation techniques (\citet{Goetghebeur2000}, \citet{Stubbendick2006}). 
\textcolor{black}{In order to impute missing values, we generated a sample from $[\bm{b}_{i},\bm{y}_{mis,i}|\hat{\bm{\gamma}},\bm{y}_{obs,i}]$. Details about the sample generation can be found in Appendix (\Cref{imputesamp}). The imputed missing values helped in computing parameter estimates and variances as in the case of complete data.}
Repeating the procedure $m_0$ times, the final variance estimates were taken to be: (mean of the imputed variances)+$(1+1/m_{0})$(empirical variances of the imputed point estimates).
\section{Hypothesis Testing and Power Computation}\label{power}
The MCEM-LRT test is developed for testing the fixed effects of a multivariate crossover model with missing data. 
Although we are interested in testing fixed effects here, the LRT procedure may also be used to test variance components. 
The LRT statistic is $\Lambda=2 [\log L\big(\hat{\bm{\theta}}_{f u l l}\big)- \log L\big(\hat{\bm{\theta}}_{\text {reduced}}\big)],$ where $\log L\big(\hat{\bm{\theta}}_{\text {full}}\big)$
is the log-likelihood function for the full model containing all the parameters while $\log L\big(\hat{\bm{\theta}}_{\text {reduced}}\big)$ is the log-likelihood function value for the reduced model under $H_{0}$. 
Under the normality assumption and certain other regularity conditions (\citet{junshao}), when $H_0$ is true, $\Lambda$ follows a central $\chi^{2}_{d f}$ distribution, where the degree of freedom, $\mathrm{df}$, is the difference between the number of parameters in the full model and the reduced model. The null hypothesis, $H_0$, is rejected if the observed value of $\Lambda$ exceeds the $(1-\alpha)^{\text{th}}$ quantile of $\chi^{2}_{d f}$.

To test the treatment effect, period, and gene effect, we formulate the following hypotheses:
\begin{align*}
H_{0}:\tau_1=\tau_2&=\ldots=\tau_t \;\;\textit{vs.}\;\; H_{1}: \tau_1,\tau_2,\ldots,\tau_t \text{ are not all equal },\\
H_{0}:\pi_1=\pi_2&=\ldots=\pi_p \;\;\textit{vs.}\;\; H_{1}: \pi_1,\pi_2,\ldots,\pi_p \text{ are not all equal},\\
H_{0}:g_1=g_2&=\ldots=g_m \;\;\textit{vs.}\;\; H_{1}: g_1,g_2,\ldots,g_m \text{ are not all equal }.
\end{align*}

In the case of MCEM-LRT, we assume that the central limit theorem applies to the MCEM estimator. Thus, in the next section, we conduct extensive simulations to assess the EM-LRT test and its power for various parameter values and sample size combinations.

\section{Simulation Study}\label{simstudy}
We discuss a detailed simulation study to assess the performance of the MCEM estimators and the LRT test based on these estimators. 
For data generation, we assumed a crossover trial with three treatment sequences $\{ABC, BAC, CBA\}$ in three periods. Two simulation scenarios with 10 and 30 subjects assigned to each treatment sequence were considered. Four response variates were measured in each period. The model is represented as,
\begin{align*}
y_{i j k l}&=\beta_{0}+\beta_{1} \text{T}_1 +\beta_{2}  \text{T}_2+\beta_{\tau_1} \text {Trt}_{1}+\beta_{\tau_2} \text {Trt}_{2}+\beta_{r_1} \text {Res}_{1}+\beta_{r_2} \text {Res}_{2}\\&+\beta_{r_3}\text {Res}_{3}+s_{i k} 
+e_{ijkl}; \; i,j=1 (1) 3, \; l=1 (1) 4, \; k=1 (1) n_i, \label{model in regression}
\end{align*}
where $y_{i j kl}$ denotes the $l^{\text{th}}$ response value from the $k^{\text{th}}$ subject in the $j^{\text{th}}$ period of the $i^{\text{th}}$ sequence; $\beta_{0}$: an intercept; T$_1$, T$_2$: indicator variable corresponding to time/period effects; $\operatorname{Trt}_{1}$, $\operatorname{Trt}_{2}$:  indicator variable corresponding to the treatment effects and Res$_{1}$, Res$_{2}$, Res$_{3}$: indicator variables corresponding to the response variate, where Res$_{i}=1$ for the $i^{\text{th}}$ response variate and $0$ otherwise for $i=1 (1) 3$. Also, as defined before, $s_{i k}$ is the subject-specific random effect, and $e_{i j kl}$ is the error term.
\[
s_{i k} \sim N\left(0, \sigma_{s}^{2}\right), \; e_{i j kl} \sim N\left(0, \sigma_{e}^{2}\right);\;  s_{ik} \perp e_{ijkl}.
\]
In scenario 1, $n_i$ was assumed to be 10 $\forall i$.
In matrix notations,
\begin{equation*}\label{eq:sim model}
\bm{y}_{i}=\bm{ X}_{i}\bm{\beta}+\bm{Z}_{i}\bm{b}_{i}+\bm{e}_{i};\; i=1 (1) 3,
\end{equation*}
where, $\bm{e}_{i} \sim N_{120}(\bm{0},1.44 \bm{I})$, $\bm{b}_{i} \sim N_{10}(\bm{0},0.49\bm{I})$, $\bm{X}_{i}$ and $\bm{Z}_{i}$ matrices are constructed similarly as in \cref{eq:2}. The true values of the components of $\bm{\beta}$ are given in the first column of \Cref{table simulated data}. 

The missing data mechanism is referred to as MAR. We assume that there are no missing values in period 1. Also, in the 2$^{\text{nd}}$ and 3$^{\text{nd}}$ period, we assume if the first response value is available, the other responses are also available. Vice versa, if the first response value is missing in the 2$^{\text{nd}}$ and 3$^{\text{nd}}$ period, the remaining response values are also assumed to be missing. Therefore, we introduce indicator variables for the $k^{\text{th}}$ subject in the $i^{\text{th}}$ sequence $(r_{ik})$ to account for missingness in periods 2, 3. That is,  $r_{iku}=(r_{ik2},r_{ik3})$ where, 
\[ r_{iku}=\begin{cases} 
      1, & \text{first response of period}\; u\; \text{is missing for} \; k^{\text{th}}\; \\ & \text{subject receiving the}\; i^{\text{th}} \text{sequence}\\
      0, & \text{ otherwise}
   \end{cases}
\]and $i=1 (1) 3$, $k=1 (1) 10$, $u= 2, 3$. It is enough to introduce an indicator variable for only the first response in periods 2 and 3. Its status determines if the rest of the response values in that specific period are missing/available. Following \citet{diggleken}, we use a binomial model for generating the missing data indicator, i.e., $$f(r \mid \phi, y)=\prod_{i=1}^{s} \prod_{k=1}^{n_{i}}\prod_{u=2}^{3}\left[P\left(r_{i ku}=1 \mid \phi\right)\right]^{r_{i ku}}\left[\left(1-P\left(r_{i ku}=1 \mid \phi\right)\right)\right]^{1-r_{i ku}}$$
where $P(r_{iku} = 1|\phi)$ is modeled via a logistic regression involving outcomes of previous time points (since MAR is assumed). It takes the form
$$\operatorname{logit}\left(P\left(r_{i ku}=1 \mid \phi\right)\right) \equiv \log \left[\frac{P\left(r_{i ku}=1 \mid \phi\right)}{1-P\left(r_{i ku}=1 \mid \phi\right)}\right]=\phi_{0}+\phi_{1} y_{i k \overline{u-1}}+\phi_{2} y_{i k \overline{u-2}}$$ where $i=1 (1) 3$, $k=1 (1) 10$, $u= 2, 3$ and for $u=2$, $\phi_2=0$. For $u= 2, 3$, $\phi_0=0.1$,  $\phi_1=-0.41$ and for $u=3$, $\phi_2=0.1$. Using these parameter values, we get, on average, 38 percent missing outcomes. Changing the parameter values to $\phi_0=\phi_2=0.1$, $\phi_1=-0.71$, \% of missingness changes to 24 percent.

Simulations were repeated 300 times for both sets of true parameter values. \Cref{table simulated data} show the average simulation results in terms of parameter estimates, SEs, average relative bias, and p-value. Average relative bias for $\hat{\beta}_u$ was computed  as $\frac{\sum_{w=1}^{500} (\hat{\beta}_{uw}-\beta_u)}{300 \beta_u }$, where $\hat{\beta}_{uw}$ is the $u^\text{th}$ component of $\hat{\bm{\beta}}$ for the $w^\text{th}$ simulation and $\beta_u$ is the true value. A total of one hundred imputations were used to obtain the SEs. \textcolor{black}{The reported p-values represent the significance of each parameter estimate using t-tests. In Figure 4, we empirically demonstrate the asymptotic normality of MCEM estimates using density plots based on these test statistics.} 

\textcolor{black}{To handle missing data, the most common and naive ways are either to perform (i) a complete case analysis, or (ii) an available case analysis, or (iii) an imputation analysis (\citet{briggs2003}, \citet{little2002}, \citet{Schafer2002}). We compare the simulation results from the proposed estimation method described in this Section  to these naive methods. A complete case in the crossover setup means that a subject who missed an observation in any period has been totally removed. Typically, complete case analysis assumes that missing data is due to MCAR or that the observed complete cases represent random samples from the originally targeted samples (\citet{pigott2001}). \citet{little2002} suggested that when the missingness is not MCAR and the complete cases are not representative of all cases, assuming complete cases and discarding the incomplete data may result in a loss of precision and bias. } 
\textcolor{black}{For using the imputation method, one needs to obtain means or draws from the predictive distribution of the missing values. Several methods are available for generating this predictive distribution, including mean imputation, regression imputation, and hot deck imputation (\citet{briggs2003}). Imputation consists of filling in missing data to create a complete data matrix. \citet{little2002} discussed that inferences about parameters based on filled-in data might not account for the associated uncertainty. As a result, standard errors may be underestimated, leading to smaller p-values and narrower confidence intervals. In available case analysis, one works with the data available by ignoring the missing observations. Applying available case analysis to our crossover settings left us with an unbalanced crossover design that is beyond this article's scope. Hence, we ended up comparing our estimation method to results from complete case analysis and hot deck imputation method.
}

 In \Cref{table simulated data}, we report the true values, estimates, SEs, relative bias, and p-values of the parameters. Two missing data percentages have been studied, $24.4\%$ and $37.4\%$. As compared to the naive methods, the relative bias and SE values are relatively stable, with an increase in missing data percentages. Using the proposed EM technique, the gene and period effects are shown to be significant at the level of significance $0.05$, however, the treatment effects are not. Tables \ref{table simulated data} and \ref{Tab:range} compare the estimation using the proposed EM technique with complete case and imputation analysis. To impute the missing values, we used the predictive mean matching method available in the MICE package of R. \textcolor{black}{From the tables, we see that the relative biases and SEs for both the complete case and imputation method are higher than the proposed EM technique. Due to an increase in bias and SEs, the corresponding p-values are unable to correctly conclude the significance of various effects.}  
 
 \textcolor{black}{In the second scenario $n_i$ was assumed to be 30 $\forall i$. Table \ref{table simulated datan30} of the Appendix (\Cref{secsim}) contains results of this simulation study and shows further improvements in both bias and SEs of the proposed EM technique due to increasing the number of subjects from 10 to 30 as compared to \Cref{table simulated data} (in which there are ten subjects in each sequence). }
 
\begin{sidewaystable}[!htp]
\scriptsize
\begin{center}
\caption{\textcolor{black}{Simulation results based on 300 samples with ten subjects per sequence for the parameter settings $\phi_0=\phi_2=0.1, \phi_1=-0.41$, $\phi_0=\phi_2=0.1, \phi_1=-0.71$ respectively: maximum likelihood estimates, SEs, relative bias and p-values for proposed analysis, complete cases and imputation analysis. True parameter values are given in parentheses.}}
\label{table simulated data}
\begin{tabular}{llllrlllrlllr} \hline 
                  & \multicolumn{12}{c}{$\phi_0=\phi_2=0.1, \phi_1=-0.71$, the overall missing percentage is 24.4}                                 \\ \hline 
                  & \multicolumn{4}{c}{Proposed analysis}   & \multicolumn{4}{c}{Complete cases}      & \multicolumn{4}{c}{Imputation analysis} \\ \hline
Parameter         & Estimate & SE     & Rel\_bias & p-value & Estimate & SE     & Rel\_bias & p-value & Estimate & SE     & Rel\_bias & p-value \\ \hline 
Intercept   (2.5) & 2.4875   & 0.1675 & -0.0050   & $\le0.0001$  & 2.5491   & 0.3247 & 0.0196    & $\le0.0001$  & 2.6383   & 0.2698 & 0.0553    & $\le0.0001$  \\
Period$_1$ (0.4)    & 0.3981   & 0.1627 & -0.0047   & 0.0149  & 0.4347   & 0.3330 & 0.0867    & 0.1929  & 0.3079   & 0.2776 & -0.2304   & 0.2683  \\
Period$_2$   (1.06) & 1.0407   & 0.1996 & -0.0182   & $\le0.0001$ & 1.1094   & 0.4098 & 0.0466    & 0.0072  & 0.9308   & 0.3400 & -0.1218   & 0.0065  \\
Trt$_1$ (0.26)      & 0.2536   & 0.1627 & -0.0247   & 0.1201  & 0.2650   & 0.3330 & 0.0191    & 0.4269  & 0.2359   & 0.2776 & -0.0929   & 0.3962  \\
Trt$_2$ (0.32)      & 0.3054   & 0.1992 & -0.0456   & 0.1262  & 0.3136   & 0.4098 & -0.0199   & 0.4448  & 0.2858   & 0.3400 & -0.1069   & 0.4012  \\
Res$_1$ (0.5)        & 0.5000   & 0.1704 & -0.0001   & 0.0036  & 0.6008   & 0.3508 & 0.2016    & 0.0880  & 0.4667   & 0.2926 & -0.0666   & 0.1117  \\
Res$_2$ (0.7)        & 0.7058   & 0.1704 & 0.0082    &$\le0.0001$ & 0.7271   & 0.3508 & 0.0387    & 0.0392  & 0.6552   & 0.2926 & -0.0640   & 0.0258  \\
Res$_3$ (0.6)        & 0.6070   & 0.1704 & 0.0117    & 0.0004  & 0.6234   & 0.3508 & 0.0390    & 0.0767  & 0.5653   & 0.2926 & -0.0578   & 0.0542  \\
$\sigma_e^2$  (1.44)     & 1.3984   & 0.1038 & -0.0289   & $\le0.0001$  & 1.4557   & 0.3588 & 0.0109    & 0.0001  & 1.3747   & 0.2879 & -0.0453   & $\le0.0001$  \\
$\sigma_s^2$ (0.49)     & 0.4759   & 0.1343 & -0.0288   & 0.0004  & 0.4808   & 0.2055 & -0.0188   & 0.0200  & 0.4030   & 0.2446 & -0.1775   & 0.1003  \\ \hline 
                  & \multicolumn{12}{c}{ $\phi_0=\phi_2=0.1, \phi_1=-0.41$, the overall missing percentage is 37.4}                               \\ \hline 
                  & \multicolumn{4}{c}{Proposed analysis}   & \multicolumn{4}{c}{Complete cases}      & \multicolumn{4}{c}{Imputation analysis} \\ \hline
Parameter         & Estimate & SE     & Rel\_bias & p-value & Estimate & SE     & Rel\_bias & p-value & Estimate & SE     & Rel\_bias & p-value \\ 
 \hline
Intercept   (2.5) & 2.5086   & 0.1642 & 0.0034    & $\le0.0001$  & 2.5525   & 0.3657 & 0.0210    & $\le0.0001$ & 2.7101   & 0.2668 & 0.0840    & $\le0.0001$  \\
Period$_1$ (0.4)    & 0.3976   & 0.1599 & -0.0060   & 0.0134  & 0.4401   & 0.3843 & 0.1002    & 0.2529  & 0.2714   & 0.2745 & -0.3215   & 0.3234  \\
Period$_2$   (1.06) & 1.0176   & 0.1960 & -0.0400   & $\le0.0001$  & 1.1039   & 0.4715 & 0.0414    & 0.0198  & 0.8226   & 0.3362 & -0.2240   & 0.0149  \\
Trt$_1$ (0.26)      & 0.2302   & 0.1599 & -0.1148   & 0.1510  & 0.2305   & 0.3843 & -0.1134   & 0.5490  & 0.2211   & 0.2745 & -0.1495   & 0.4210  \\
Trt$_2$ (0.32)      & 0.2903   & 0.1960 & -0.0927   & 0.1395  & 0.2867   & 0.4715 & -0.1040   & 0.5435  & 0.2726   & 0.3362 & -0.1481   & 0.4180  \\
Res$_1$ (0.5)        & 0.4887   & 0.1663 & -0.0225   & 0.0035  & 0.5817   & 0.3966 & 0.1633    & 0.1434  & 0.4343   & 0.2893 & -0.1313   & 0.1342  \\
Res$_2$ (0.7)        & 0.6880   & 0.1663 & -0.0172   & $\le0.0001$  & 0.7098   & 0.3966 & 0.0141    & 0.0744  & 0.6114   & 0.2893 & -0.1266   & 0.0353  \\
Res$_3$ (0.6)        & 0.6028   & 0.1663 & 0.0047    & 0.0003  & 0.6356   & 0.3966 & 0.0593    & 0.1099  & 0.5349   & 0.2893 & -0.1085   & 0.0653  \\
$\sigma_e^2$ (1.44)     & 1.3962   & 0.1002 & -0.0304   & $\le0.0001$  & 1.5078   & 0.4166 & 0.0471    & 0.0003  & 1.3414   & 0.2814 & -0.0685   & $\le0.0001$  \\
$\sigma_s^2$ (0.49)     & 0.4624   & 0.1278 & -0.0563   & 0.0003  & 0.4613   & 0.1462 & -0.0585   & 0.0017  & 0.3525   & 0.2090 & -0.2806   & 0.0926 \\ \hline 
\end{tabular}
\end{center}
\end{sidewaystable}
\begin{table}[!htp]
\centering
\scriptsize
 \caption{\textcolor{black}{Range of SEs and relative biases under proposed analysis, complete cases, and imputation analysis: 10 subjects per sequence with the parameter settings $\phi_0=\phi_2=0.1, \phi_1=-0.71$ and $\phi_0=\phi_2=0.1, \phi_1=-0.41$.}}
 \label{Tab:range}
\begin{tabular}{llll}\hline
              & \multicolumn{3}{c}{For  $\phi_0=\phi_2=0.1, \phi_1=-0.71$} \\  \hline
              & Proposed analysis  & Complete cases  & Imputation analysis \\ \hline 
SE            & (0.10, 0.20)       & (0.21, 0.41)    & (0.24, 0.34)        \\  
Relative Bias & (-0.05, 0.01)      & (-0.02, 0.20)   & (-0.23, 0.06)       \\ \hline 
              & \multicolumn{3}{c}{For  $\phi_0=\phi_2=0.1, \phi_1=-0.41$} \\ \hline 
SE            & (0.10, 0.20)       & (0.15, 0.47)    & (0.21, 0.34)        \\
Relative Bias & (-0.11, 0.00)      & (-0.11, 0.16)   & (-0.32, 0.08)  \\ \hline     
\end{tabular}
\end{table}

\begin{figure}[htp]
\centering
  \includegraphics[width=12cm]{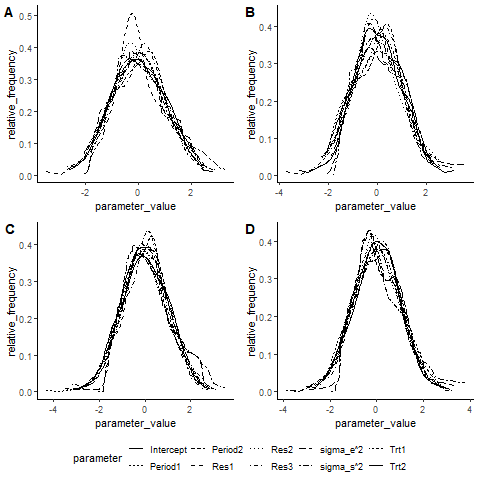}
  \caption{ Empirical PDF of standardized MCEM estimates; Plots (A, B) corresponds to cases with 100 simulated data sets when $\phi_0=\phi_2=0.1$ and $\phi_1$ is $-0.71$ and $-0.41$, respectively. Plots (C, D) corresponds to cases with 300 simulated data sets when $\phi_0=\phi_2=0.1$ and $\phi_1$ is $-0.71$ and $-0.41$, respectively. }
  \label{fig:asy norm}
 \end{figure}

Empirical probability density curves for each standardized estimator were compared with standard normal density curves to verify asymptotic normality. \Cref{fig:asy norm} displays an empirical PDF of the standardized MCEM estimates with the number of simulations. According to \Cref{fig:asy norm}, the MCEM estimators (standardized one) are closer to normality as the number of simulations increases. 

Further, we tested the following hypotheses for the simulation setting with $\phi_0=\phi_2=0.1$ and $\phi_1=-0.41$ using the MCEM-LRT procedure detailed in \Cref{power} and computed the  power of the following tests: \begin{itemize}
\item[(i)] Tests for gene effects: $H_{0}: \beta_{r_1}=\beta_{r_2}=\beta_{r_3}=0$ versus $H_1$: at least one $\beta_{r_i}\neq 0$ for $i=1 (1) 3$

\item[(ii)] Pairwise tests of gene effects
\item[(iii)] To test for the treatment effect we used, $H_{0}: \beta_{\tau_1}=\beta_{\tau_2}$ vs $H_{1}: \beta_{\tau_1} \neq \beta_{\tau_2}$ 
\end{itemize}

For empirical power computations, we list the steps:
\begin{itemize}
\item[(a)] For data simulation, we used the true values of the parameters from \Cref{table simulated data}. The proposed MCEM algorithm was used to find the estimates of the unknown parameters and variance components, $\bm{\theta}$, and denoted as $\hat{\bm{\theta}}_{full}$. 
\item[(b)] To obtain the restricted estimates, we maximized the Q- function under $H_{0}$ and computed $\hat{\bm{\theta}}_{reduced}$.
\item[(c)] Using $\hat{\bm{\theta}}_{full}$ and  $\hat{\bm{\theta}}_{reduced}$, the test statistic  $\Lambda=2 (Q(\hat{\bm{\theta}})- Q(\hat{\bm{\theta}}_{0}))$ was determined and compared with  $C=\chi^2_{(1-\alpha, d f)}$, where df is the difference between the number of parameters in the full and reduced model. 
\item[(d)] Repeat steps (a) - (d) 1000 times.
 \item[(e)]  Empirical power is calculated as the fraction of times $H_0$ has been rejected.
\end{itemize}

\begin{figure}[H]
\centering
  \includegraphics[width=8cm]{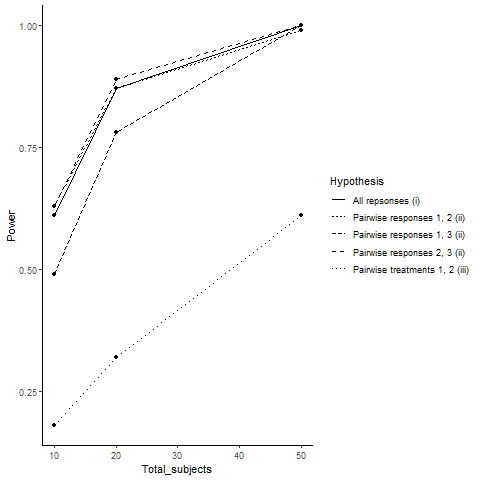}
  \caption{ Power values  of MCEM LRTs: The points plotted indicate the empirical proportion of tests (by use of a nominal level $\alpha$ = 0.05) that rejected $H_0$ among 1000 simulated data sets. }
  \label{fig:power plots gene}
 \end{figure}

\Cref{fig:power plots gene} shows the variation in the power function to sample size. We note that power increases with sample size (or the number of total subjects) from these plots. For hypotheses (i) and (ii), the power is $0.50 - 0.61$ when the total subjects are ten and increases to 1 when we increase the total subjects to 50. 
Similarly, the power for hypothesis (iii) is 0.18 when the total number of subjects is ten, and it increases to 0.61 when we increase the subjects to 50. We also observe that power values in \Cref{fig:power plots gene} (hypothesis (iii)) are low due to the small effect size (difference between $\beta_{\tau_1}$ and $\beta_{\tau_2}$ is small). 
 
\section{Gene Case Study Results}\label{reldat}
 Following the  exploratory analysis discussed in \Cref{case study}, we fitted the following normal cross-over model:
\begin{align*}
y_{i j k l}&=\beta_{0}+\beta_{1} \text {Period}_{2}+\beta_{2} \text {Period}_{3}+\beta_{3} \mathrm{Trt}_{2}+\beta_{4} \mathrm{Trt}_{3}+\beta_{5} \text {Gene}_{2}\\&+\beta_{6} \text {Gene}_{3}+\beta_{7} \text {Gene}_{4}
+\beta_{8} \text {Gene}_{5}+\beta_{9} \text {Gene}_{6}+\beta_{10} \text {Gene}_{7}\\&+\beta_{11} \text {Gene}_{8}+\beta_{12} \text {Gene}_{9}+\beta_{13} \text {Gene}_{10}+s_{i k}
+e_{i j k l} \numberthis \label{eq:11}
\end{align*}
where for the real data $i,j=1 (1) 3;\, l=1(1)10$ and $k=1(1)n_i$, where $n_1=n_2=6$ and $n_3=5$. The random effects $s_{i k}$ and $e_{i j k l}$ were assumed to be independent normal variates with  variances $\sigma_{s}^{2}$ and $\sigma_{e}^{2}$, respectively. Indicator variables were used to represent the period, treatment, and gene effects. Based on Little's MCAR test, as described in \Cref{case study}, we assumed the missing data mechanism was MAR. 

In Table \ref{table real data2}, we present the maximum likelihood estimates of parameters $(\bm{\beta},\sigma_{e}^2,\sigma_{s}^2)$, standard errors, and p-values from the normal crossover model presented in \cref{eq:11}. \textcolor{black}{A comparison of the proposed analysis versus the complete case and imputation method is also shown.} ANOVA estimates were used as initial values for the MCEM algorithm. One hundred imputations were used for standard error estimation. We computed AIC and BIC values along with the root mean square error (RMSE) of residuals to test the model's adequacy. 

From Table \ref{table real data2}, we observe that Gene$_2$, Gene$_3$, Gene$_4$, Gene$_6$, Gene$_7$, Gene$_8$, Gene$_9$, Gene$_{10}$ were significant at a 1\% level of significance in complete cases, imputation method, and missing case results. From the negative coefficients of the gene effects, we saw that the presence of that particular gene decreases the allergic reaction.
\textcolor{black}{Also, we note that the AIC and BIC values of the MAR responses model were much smaller as compared to the complete case and imputation analysis.} Moreover, working with log(responses) instead of the original responses further reduced the AIC and BIC values. 

\begin{figure}{H}
\centering
  \includegraphics{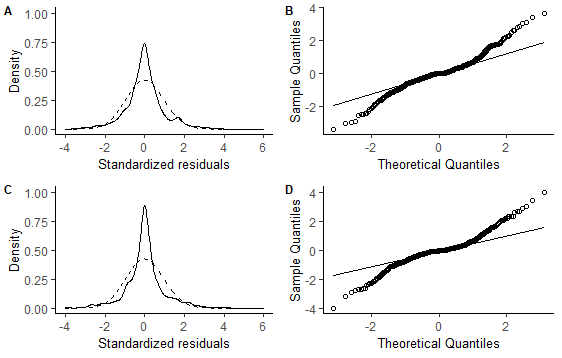}
  \caption{Density and normal Q-Q plots of conditional residuals: Plots (A, B) corresponds to the original responses. Plots (C, D) correspond to the log-transformed responses. True normal density is indicated by a dashed line.}
  \label{fig:condresid}
 \end{figure}

\textcolor{black}{
Furthermore, to assess the goodness of fit, conditional residuals were computed for the original and log-transformed models, as well as their estimated density and normal Q-Q plots. 
We can see from \Cref{fig:condresid} that there is slight evidence of a departure from normality. An alternate way of analyzing the data may be to use non-parametric methods or asymmetric families of distributions from errors, such as skew-normal or skew-t distributions. However, since the primary interest is in proposing an estimation method for handling missing responses in a multivariate crossover setup, we did not investigate these alternate approaches.}

\textcolor{black}{As can be seen from the conditional residual plots in \Cref{fig:condresid}, the original and log(response) do not differ much. However, from the AIC and BIC reported in Tables \ref{table real data2}, and \ref{table real data more missing} we see an indication that log(response) model is more appropriate. Thus based on the AIC and BIC values, we select the log response model for analysis.
Also, biomedical and clinical trial data are commonly transformed using the log transformation to reduce variability and make them conform to normality. Log transformation is included in most statistical packages due to its popularity and ease of use (\citet{Jones}).} 

We also artificially increased the number of missing responses from $10\%$ to $21\%$ and $24\%$, results are given in \Cref{table real data more missing}. We used a Bernoulli random variable for creating the artificially missing responses. In the $21\%$ case, we assumed that all other responses would also be missing if any gene had a missing response. In this way, the data set mimicked the original gene data set. However, we assumed that only some gene responses (not necessarily all) were missing in the $24\%$ missingness case. Moreover, the gene effects remained significant despite the higher missing percentage as well. We also calculated AIC, BIC, and RMSE values for responses and log(responses). \textcolor{black}{In this case, too, log transformation is more effective based on AIC and BIC values.}

\section{Computational Details}
\textcolor{black}{
For simulation and real data analyses, R programming (\citet{rr}) has been used. We used R version 4.1.0 under Windows 10 (64-bit), with an Intel core I5 processor and 4GB of RAM. The foreach function of the library doParallel has been used to parallelize the computations.}

\textcolor{black}{
In the MCEM algorithm, for a particular simulated data set, the generation of $B=2000$ samples described in \Cref{imputesamp} takes around 2 minutes. For one hundred simulated data sets using the simulation specifications described in Section 6 with ten subjects per sequence, it takes about 2.5 hours to run. We defined the convergence criterion as the difference between estimated values at $(t+1)^{\text{th}}$ iteration and $(t)^{\text{th}}$ iteration being less than $5\times 10^{-4}$. The MCEM algorithm converged in approximately 15-20 steps in a given set of simulated data.} 

\textcolor{black}{The data were analyzed using naive missing data methods using R functions. For complete cases, the `lme' function of the nlme library of R was used to estimate parameters. The imputation was carried out using the function `mice' from the library MICE. We generated the five imputed data sets by applying the predictive mean matching (PMM) method to fill in the missing values. The average of these five imputed sets was used as the final complete data set.} 

\textcolor{black}{In the context of the LR test, taking into account the computing power of the hypothesis defined in Section 6, it took about 1.5 hours to run it on ten subjects.
For the gene data, the mcar\_test function within the naniar library is used to verify the MCAR assumption. The gene data can be processed in approximately 3-4 minutes using similar settings on the R platform. These R programs are available at the hyperlink, \href{https://github.com/savitapareek/Likelihood-based-Missing-Data-Analysis-in-Crossover-Trials.git}{rprograms}.}
\begin{table}[!htp]
\scriptsize
\centering
\caption{\textcolor{black}{Analysis of gene data: comparison of proposed analyses, complete cases, and imputation methods using AIC and BIC values. The maximum likelihood estimates, SEs, and p-values were calculated using the original gene data with actual and log-transformed responses.}}
\label{table real data2}
\begin{tabular}{lrrrrrr}\hline
          & \multicolumn{3}{c}{\begin{tabular}[c]{@{}c@{}}Proposed   analysis  \\        (log(responses))\end{tabular}} & \multicolumn{3}{c}{\begin{tabular}[c]{@{}c@{}}Proposed   analysis  \\        (Original responses)\end{tabular}} \\ \hline
Parameter & Estimate                         & SE                             & p-value                                 & Estimate                          & SE                              & p-value                                   \\ \hline
Intercept & 1.2502                           & 0.0079                         & $\le$0.0001                       & 3.4898                            & 0.0174                          & $\le$0.0001                         \\
Period$_2$   & -0.0026                          & 0.0051                         & 0.5800                                  & -0.0024                           & 0.0111                          & 0.7300                                    \\
Period$_3$ & 0.0042                           & 0.0051                         & 0.3500                                  & 0.0098                            & 0.0112                          & 0.3900                                    \\
Trt$_2$     & -0.0028                          & 0.0051                         & 0.5700                                  & -0.0053                           & 0.0111                          & 0.5800                                    \\
Trt$_3$      & 0.0024                           & 0.0051                         & 0.6100                                  & 0.0056                            & 0.0111                          & 0.5800                                    \\
Gene$_2$     & -0.3161                          & 0.0092                         & $\le$0.0001                       & -0.9372                           & 0.0202                          & $\le$0.0001                         \\
Gene$_3$     & -0.6063                          & 0.0092                         & $\le$0.0001                       & -1.5855                           & 0.0202                          & $\le$0.0001                         \\
Gene$_4$     & -0.0982                          & 0.0092                         & $\le$0.0001                       & -0.3258                           & 0.0202                          & $\le$0.0001                         \\
Gene$_5$     & -0.0103                          & 0.0092                         & 0.2800                                  & -0.0357                           & 0.0202                          & 0.0800                                    \\
Gene$_6$     & -0.4512                          & 0.0092                         & $\le$0.0001                       & -1.2648                           & 0.0202                          & $\le$0.0001                         \\
Gene$_7$     & -0.5487                          & 0.0092                         & $\le$0.0001                       & -1.4704                           & 0.0202                          & $\le$0.0001                         \\
Gene$_8$     & -0.8659                          & 0.0092                         & $\le$0.0001                       & -2.0186                           & 0.0202                          & $\le$0.0001                         \\
Gene$_9$     & -0.2669                          & 0.0092                         & $\le$0.0001                       & -0.8163                           & 0.0202                          & $\le$0.0001                         \\
Gene$_{10}$    & -0.2019                          & 0.0092                         & $\le$0.0001                       & -0.6344                           & 0.0202                          &$\le$0.0001                         \\
$\sigma_e^2$   & 0.0024                           & 0.0001                         & -                                       & 0.0115                            & 0.0007                          & -                                         \\
$\sigma_s^2$   & 0.0001                           & 0.00004                        & -                                       & 0.0007                            & 0.0003                          & -                                         \\ \hline
AIC       & -2686.08                       &                                &                                         & -1853.11                          &                                 &                                           \\
BIC       & -2618.33                        &                                &                                         & -1785.36                          &                                 &                                           \\
RMSE      & 0.0460                           &                                &                                         & 0.1006                            &                                 &                                           \\ \hline
          & \multicolumn{3}{c}{Complete cases}                                                                           & \multicolumn{3}{c}{Imputation analysis}                                                                         \\  \hline
Parameter & Estimate                         & SE                             & p-value                                 & Estimate                          & SE                              & p-value                                   \\ \hline
Intercept & 3.4900                           & 0.0220                         & $\le$0.0001                       & 3.3880                           & 0.0388                          & $\le$0.0001                         \\
Period$_2$   & 0.0010                           & 0.0138                         & 0.9629                                  & -0.0020                           & 0.0253                          & 0.9354                                    \\
Period$_3$  & 0.0120                           & 0.0138                         & 0.4047                                  & -0.0400                           & 0.0253                          & 0.1187                                    \\
Trt$_2$      & -0.0050                          & 0.0138                         & 0.6972                                  & 0.0010                            & 0.0253                          & 0.9580                                    \\
Trt$_3$      & 0.0080                           & 0.0138                         & 0.5670                                  & 0.0110                            & 0.0253                          & 0.6579                                    \\
Gene$_2$     & -0.9280                          & 0.0252                         & $\le$0.0001                       & -0.8300                           & 0.0460                          & $\le$0.0001                         \\
Gene$_3$     & -1.5860                          & 0.0252                         & $\le$0.0001                       & -1.4270                           & 0.0460                          & $\le$0.0001                         \\
Gene$_4$     & -0.3270                          & 0.0252                         & $\le$0.0001                       & -0.2880                           & 0.0460                          & $\le$0.0001                                    \\
Gene$_5$     & -0.0380                          & 0.0252                         & 0.1396                                  & -0.0130                           & 0.0460                         & 0.7737                                    \\
Gene$_6$     & -1.2650                          & 0.0252                         & $\le$0.0001                       & -1.1210                           & 0.0460                          & $\le$0.0001                         \\
Gene$_7$     & -1.4750                          & 0.0252                         & $\le$0.0001                       & -1.3170                           & 0.0643                          & $\le$0.0001                         \\
Gene$_8$     & -2.0330                          & 0.0252                         & $\le$0.0001                       & -1.8220                           & 0.0460                          & $\le$0.0001                         \\
Gene$_9$     & -0.8210                          & 0.0252                         & $\le$0.0001                       & -0.7380                           & 0.0460                          & $\le$0.0001                         \\
Gene$_{10}$    & -0.6480                          & 0.0252                         & $\le$0.0001                       & -0.5520                          & 0.0460                          & $\le$0.0001                         \\
$\sigma_e^2$   & 0.0115                           & 0.0009                         & -                                       & 0.0540                            & 0.0034                          & -                                         \\
$\sigma_s^2$   & 0.0005                           & 0.0004                         & -                                       & 0.0000                            & 0.0006                          & -                                         \\ \hline
AIC       & -1303.27                       &                                &                                         & -1832.63                         &                                 &                                           \\
BIC       & -1241.09                        &                                &                                         & -1764.88                          &                                 &                                           \\
RMSE      & 0.1070                           &                                &                                         & 0.2324                            &                                 &   \\ \hline                                       
\end{tabular}
\end{table}
\begin{table}[!htp]
\scriptsize
\centering
\caption{\textcolor{black}{Analysis of gene data with artificially increased missingness: maximum likelihood estimates, SEs, and p-value for missing at random responses on original and logarithmic scales.}}
\label{table real data more missing}
\begin{tabular}{lrrrrrr}\hline
 & \multicolumn{6}{c}{21\% missing}\\ \hline
          & \multicolumn{3}{c}{\begin{tabular}[c]{@{}c@{}}Proposed   analysis  \\        (log(responses))\end{tabular}} & \multicolumn{3}{c}{\begin{tabular}[c]{@{}c@{}}Proposed   analysis  \\        (Original responses)\end{tabular}} \\ \hline
Parameter & Estimate                         & SE                             & p-value                                 & Estimate                          & SE                              & p-value                                   \\ \hline
Intercept & 1.2503      & 0.0070   & $\le$0.0001    & 3.4895       & 0.0157     & $\le$0.0001    \\
Period$_2$   & -0.0027     & 0.0044   & 0.3900    & -0.0026      & 0.0098     & 0.4800     \\
Period$_3$  & 0.0054      & 0.0044   & 0.2100    & 0.0119       & 0.0101     & 0.2800     \\
Trt$_2$      & -0.0005     & 0.0044   & 0.5700    & -0.0005      & 0.0100     & 0.5300     \\
Trt$_3$      & 0.0005      & 0.0044   & 0.5300    & 0.0031       & 0.0100     & 0.7400     \\
Gene$_2$     & -0.3119     & 0.0080   & $\le$0.0001   & -0.9290      & 0.0179     & $\le$0.0001     \\
Gene$_3$     & -0.6104     & 0.0080   & $\le$0.0001    & -1.5932      & 0.0179     & $\le$0.0001    \\
Gene$_4$     & -0.0989     & 0.0080   & $\le$0.0001    & -0.3284      & 0.0179     & $\le$0.0001    \\
Gene$_5$     & -0.0106     & 0.0080   & 0.1900    & -0.0367      & 0.0179     & 0.0600     \\
Gene$_6$     & -0.4529     & 0.0080   & $\le$0.0001   & -1.2690      & 0.0179     & $\le$0.0001     \\
Gene$_7$     & -0.5508     & 0.0080   & $\le$0.0001    & -1.4747      & 0.0179     & $\le$0.0001    \\
Gene$_8$     & -0.8654     & 0.0080   & $\le$0.0001    & -2.0188      & 0.0179     & $\le$0.0001    \\
Gene$_9$     & -0.2675     & 0.0080   & $\le$0.0001    & -0.8175      & 0.0179     & $\le$0.0001     \\
Gene$_{10}$    & -0.1997     & 0.0080   & $\le$0.0001   & -0.6284      & 0.0179     & $\le$0.0001     \\
$\sigma_e^2$   & 0.0021      & 0.0001   & -         & 0.0105       & 0.0005     & -          \\
$\sigma_s^2$   & 0.0001      & 0.0000   & -         & 0.0007       & 0.0003     & -          \\ \hline
AIC       & -2758.21    &          &           & -1907.18     &            &            \\
BIC       & -2690.46    &          &           & -1839.43     &            &            \\
RMSE      & 0.0397      &          &           & 0.0884       &            &            \\ \hline
          & \multicolumn{6}{c}{24\% missing}                                            \\ \hline
          & \multicolumn{3}{c}{log(responses)} & \multicolumn{3}{c}{Original responses} \\ \hline
Intercept & 1.2499      & 0.0076   & $\le$0.0001    & 3.4889       & 0.0166     & $\le$0.0001     \\
Period$_2$   & -0.0050     & 0.0048   & 0.1800    & -0.0084      & 0.0105     & 0.2400     \\
Period$_3$ & 0.0022      & 0.0048   & 0.3800    & 0.0045       & 0.0105     & 0.4500     \\
Trt$_2$      & -0.0021     & 0.0048   & 0.4200    & -0.0037      & 0.0105     & 0.4500     \\
Trt$_3$      & 0.0047      & 0.0048   & 0.2500    & 0.0104       & 0.0105     & 0.2900     \\
Gene$_2$     & -0.3130     & 0.0088   & $\le$0.0001    & -0.9286      & 0.0190     & $\le$0.0001    \\
Gene$_3$     & -0.6051     & 0.0088   & $\le$0.0001    & -1.5828      & 0.0190     & $\le$0.0001   \\
Gene$_4$     & -0.0985     & 0.0088   & $\le$0.0001    & -0.3270      & 0.0190     & $\le$0.0001     \\
Gene$_5$     & -0.0098     & 0.0088   & 0.2100    & -0.0349      & 0.0190     & 0.0800     \\
Gene$_6$     & -0.4511     & 0.0088   & $\le$0.0001    & -1.2645      & 0.0190     & $\le$0.0001   \\
Gene$_7$    & -0.5495     & 0.0088   & $\le$0.0001    & -1.4711      & 0.0190     & $\le$0.0001   \\
Gene$_8$     & -0.8665     & 0.0088   & $\le$0.0001    & -2.0180      & 0.0190     & $\le$0.0001  \\
Gene$_9$     & -0.2683     & 0.0088   & $\le$0.0001    & -0.8198      & 0.0190     & $\le$0.0001    \\
Gene$_{10}$    & -0.2027     & 0.0088   & $\le$0.0001    & -0.6369      & 0.0190     & $\le$0.0001    \\
$\sigma_e^2$   & 0.0025      & 0.0001   & -         & 0.0119       & 0.0006     & -          \\
$\sigma_s^2$   & 0.0001      & 0.0000   & -         & 0.0006       & 0.0003     & -          \\ \hline
AIC       & -2652.14    &          &           & -1830.59     &            &            \\
BIC       & -2584.39    &          &           & -1762.84     &            &            \\
RMSE      & 0.0436      &          &           & 0.0941       &            &     \\ \hline      
\end{tabular} 
\end{table}

\section{Concluding Remarks} In this paper, we studied a statistical model to analyze incomplete data in a multivariate crossover setup. \textcolor{black}{\Cref{sec mle missing} presents a maximum likelihood estimation for crossover design with multiple responses and missing observations. It is a specific situation of the linear mixed model where crossover design is considered. Several modifications to the mixed model setup are required to accommodate data from a cross-over trial  due to its unique attributes.
To the best of our knowledge, this is the first study to introduce parameter estimation for crossover designs measuring multiple and missing responses. The presence of multiple responses in each period complicates the covariance structure of the model considered.  
In crossover studies, the response variable can have missing values in either a monotone or non-monotone pattern. Due to a lack of information about the causes of missingness in the allergen challenge, we assume the missing mechanism to be MAR. The MCEM algorithm for maximum likelihood estimation is used for fitting the data.}

\textcolor{black}{Though we did not discuss NMAR type of missingness in the context of multivariate crossover trials, the non-response may very well be linked with an unobserved variable as in most cancer trials.  As a future direction, we would like to investigate the effect of including a parametric model for the missingness mechanism and see the effect on estimation in such a multivariate situation. We also would like to investigate the consequences of using a skewed normal model to fit the data from the nasal allergen case study as departures from normality were detected in the gene data. }

\vspace*{0.6pc}
\noindent {\bf{Acknowledgement}}
We thank Dr. Atanu Bhattacharjee, Tata Memorial Center, Mumbai, India, for his assistance in obtaining the gene data set.
\vspace*{0.6pc}
\noindent {\bf{Conflict of Interest}}
\noindent {{The authors have declared no conflict of interest.}}
\bibliographystyle{plainnat}
\bibliography{library}

\begin{thebibliography}{36}
\providecommand{\natexlab}[1]{#1}
\providecommand{\url}[1]{\texttt{#1}}
\expandafter\ifx\csname urlstyle\endcsname\relax
  \providecommand{\doi}[1]{doi: #1}\else
  \providecommand{\doi}{doi: \begingroup \urlstyle{rm}\Url}\fi

\bibitem[Basu and Santra(2010)]{Basu2010}
Sanjib Basu and Sourav Santra.
\newblock {A joint model for incomplete data in crossover trials}.
\newblock \emph{Journal of Statistical Planning and Inference}, 140\penalty0
  (10):\penalty0 2839--2845, 2010.

\bibitem[Briggs et~al.(2003)Briggs, Clark, Wolstenholme, and
  Clarke]{briggs2003}
Andrew Briggs, Taane Clark, Jane Wolstenholme, and Philip Clarke.
\newblock {Missing....presumed at random: Cost-analysis of incomplete data}.
\newblock \emph{Health Economics}, 12\penalty0 (5):\penalty0 377--392, 2003.

\bibitem[Chan and Ledolter(1995)]{chan1995}
K.S. Chan and Johannes Ledolter.
\newblock {Monte Carlo EM estimation for time series models involving counts}.
\newblock \emph{Journal of the American Statistical Association}, 90\penalty0
  (429):\penalty0 242--252, 1995.

\bibitem[Chinchilli and Esinhart(1996)]{chin1996}
Vernon~M. Chinchilli and James~D. Esinhart.
\newblock {Design and analysis of intra-subject variability in cross-over
  experiments}.
\newblock \emph{Statistics in Medicine}, 15:\penalty0 1619--1634, 1996.

\bibitem[Chow and Shao(1997)]{Chow1997}
Shein-chung Chow and Jun Shao.
\newblock {Statistical methods for two-sequence three-period cross-over designs
  with incomplete data}.
\newblock \emph{Statistics in Medicine}, 16:\penalty0 1031--1039, 1997.

\bibitem[Clough and Barrett(2016)]{Clough2016}
Emily Clough and Tanya Barrett.
\newblock {The Gene Expression Omnibus database}.
\newblock \emph{Methods in Molecular Biology}, 1418:\penalty0 93--110, 2016.

\bibitem[Dempster et~al.(1977)Dempster, Laird, and Rubin]{DEMP1977}
A.P. Dempster, N.M. Laird, and D.B. Rubin.
\newblock Maximum likelihood from incomplete data via the {EM} algorithm.
\newblock \emph{Journal of the Royal Statistical Society: Series B
  (Methodological)}, 39:\penalty0 1--38, 1977.

\bibitem[Diggle and Kenward(1994)]{diggleken}
P.~Diggle and M.~G. Kenward.
\newblock {Informative Drop-Out in Longitudinal Data Analysis}.
\newblock \emph{Journal of the Royal Statistical Society. Series C (Applied
  Statistics)}, 43\penalty0 (1):\penalty0 49--93, 1994.

\bibitem[Fort and Moulines(2003)]{Fort2003}
Gersende Fort and {\'E}ric Moulines.
\newblock Convergence of the monte carlo expectation maximization for curved
  exponential families.
\newblock \emph{Annals of Statistics}, 31:\penalty0 1220--1259, 2003.

\bibitem[Gelfand and Smith(1990)]{gelfrad}
Alan~E. Gelfand and Adrian~F.M. Smith.
\newblock {Sampling-based approaches to calculating marginal densities}.
\newblock \emph{Journal of the American Statistical Association}, 85\penalty0
  (410):\penalty0 398--409, 1990.

\bibitem[Geman and Geman(1984)]{geman1984}
Stuart Geman and Donald Geman.
\newblock {Stochastic Relaxation, Gibbs Distributions, and the Bayesian
  Restoration of Images}.
\newblock \emph{IEEE Transactions on Pattern Analysis and Machine
  Intelligence}, PAMI-6\penalty0 (6):\penalty0 721--741, 1984.

\bibitem[Goetghebeur and Ryan(2000)]{Goetghebeur2000}
Els Goetghebeur and Louise Ryan.
\newblock {Semiparametric regression analysis of interval-censored data}.
\newblock \emph{Biometrics}, 56\penalty0 (4):\penalty0 1139--1144, 2000.

\bibitem[Grender and Johnson(1993)]{Grender1993}
Julie~M. Grender and William~D. Johnson.
\newblock {Analysis of crossover designs with multivariate response}.
\newblock \emph{Statistics in Medicine}, 12\penalty0 (1):\penalty0 69--89,
  1993.

\bibitem[Ho et~al.(2012)Ho, Matthews, Henderson, Farewell, and Rodgers]{Ho2012}
Weang~K. Ho, John~N.S. Matthews, Robin Henderson, Daniel Farewell, and
  Lauren~R. Rodgers.
\newblock {Dropouts in the AB/BA crossover design}.
\newblock \emph{Statistics in Medicine}, 31\penalty0 (16):\penalty0 1675--1687,
  2012.

\bibitem[Ibrahim and Molenberghs(2009)]{Ibrahim2009}
J.G. Ibrahim and Geert Molenberghs.
\newblock {Missing data methods in longitudinal studies: A review}.
\newblock \emph{Test}, 18\penalty0 (1):\penalty0 1--43, 2009.

\bibitem[Jones and Kenward(2003)]{Jones}
Byron Jones and Michael~G. Kenward.
\newblock \emph{{Design and Analysis of Cross-Over Trials}}.
\newblock Chapman {\&} Hall/CRC, second edition, 2003.

\bibitem[Leaker et~al.(2016)Leaker, Malkov, Mogg, Ruddy, Nicholson, Tan,
  Tribouley, and Chen]{Leaker2016}
B.R. Leaker, V.A. Malkov, R.~Mogg, M.K. Ruddy, G.C. Nicholson, A.J. Tan,
  C.~Tribouley, and G.~Chen.
\newblock {The nasal mucosal late allergic reaction to grass pollen involves
  type 2 inflammation ( IL-5 and IL-13 ), the inflammasome ( IL-1 b ), and
  complement}.
\newblock \emph{Nature}, 10\penalty0 (2):\penalty0 408--420, 2016.

\bibitem[Levine and Casella(2001)]{casella2001}
Richard~A. Levine and George Casella.
\newblock {Implementations of the Monte Carlo EM Algorithm}.
\newblock \emph{Journal of Computational and Graphical Statistics}, 10\penalty0
  (3):\penalty0 422--439, 2001.

\bibitem[Little(1988)]{littletest}
R.J.A. Little.
\newblock {A test of missing completely at random for multivariate data with
  missing values}.
\newblock \emph{Journal of the American Statistical Association}, 83\penalty0
  (404):\penalty0 1198--1202, 1988.

\bibitem[Little(1995)]{Little1995}
R.J.A. Little.
\newblock {Modeling the drop-out mechanism in repeated-measures studies}.
\newblock \emph{Journal of the American Statistical Association}, 90\penalty0
  (431):\penalty0 1112--1121, 1995.

\bibitem[Little and Rubin(2002)]{little2002}
R.J.A. Little and D.B. Rubin.
\newblock \emph{{Statistical Analysis with Missing Data}}.
\newblock John Wiley \& Sons, Inc., second edition, 2002.

\bibitem[Matthews and Henderson(2013)]{Matthews2013}
John~N.S. Matthews and Robin Henderson.
\newblock {Two-period, two-treatment crossover designs subject to non-ignorable
  missing data}.
\newblock \emph{Biostatistics}, 14\penalty0 (4):\penalty0 626--638, 2013.

\bibitem[Patel(1985)]{Patel1985}
H.~I. Patel.
\newblock {Analysis of incomplete data in a two-period crossover design with
  reference to clinical trials}.
\newblock \emph{Biometrika}, 72\penalty0 (2):\penalty0 411--418, 1985.

\bibitem[Pigott(2001)]{pigott2001}
Therese~D. Pigott.
\newblock {A review of methods for missing data}.
\newblock \emph{International Journal of Phytoremediation}, 21\penalty0
  (1):\penalty0 353--383, 2001.

\bibitem[Putt and Chinchilli(1999)]{Putt1999}
Mary Putt and Vernon~M. Chinchilli.
\newblock {A mixed effects model for the analysis of repeated measures
  cross-over studies}.
\newblock \emph{Statistics in Medicine}, 18\penalty0 (22):\penalty0 3037--3058,
  1999.

\bibitem[{R Core Team}(2022)]{rr}
{R Core Team}.
\newblock \emph{R: A Language and Environment for Statistical Computing}.
\newblock R Foundation for Statistical Computing, Vienna, Austria, 2022.
\newblock URL \url{https://www.R-project.org/}.

\bibitem[Richardson and Flack(1992)]{Richardson1992}
Barbra~A. Richardson and Virginia~F. Flack.
\newblock {The analysis of incomplete data in the three-period two-treatment
  crossover design for clinical trials}.
\newblock \emph{Controlled Clinical Trials}, 13\penalty0 (5):\penalty0 381,
  1992.

\bibitem[Rosenkranz(2015)]{Rosenkranz2015}
Gerd~K. Rosenkranz.
\newblock {Analysis of cross-over studies with missing data}.
\newblock \emph{Statistical Methods in Medical Research}, 24\penalty0
  (4):\penalty0 420--433, 2015.

\bibitem[Rubin(1976)]{rubin}
D.B. Rubin.
\newblock {Inference and Missing Data}.
\newblock \emph{Biometrika}, 63\penalty0 (3):\penalty0 581--592, 1976.

\bibitem[Schafer and Graham(2002)]{Schafer2002}
Joseph~L. Schafer and John~W. Graham.
\newblock Missing data: our view of the state of the art.
\newblock \emph{Psychological methods}, 7:\penalty0 147--77, 2002.

\bibitem[Senn(2002)]{senn1993}
Stephen Senn.
\newblock \emph{{Cross-Over Trials in Clinical Research.}}
\newblock John Wiley {\&} Sons, Ltd., 2002.

\bibitem[Shao(2003)]{junshao}
Jun Shao.
\newblock \emph{{Mathematical Statistics}}.
\newblock Springer International Publishing, second edition, 2003.

\bibitem[Stubbendick and Ibrahim(2003)]{Stubbendick2003}
A.L. Stubbendick and J.G. Ibrahim.
\newblock {Maximum Likelihood Methods for Nonignorable Missing Responses and
  Covariates in Random Effects Models}.
\newblock \emph{Biometrics}, 59\penalty0 (4):\penalty0 1140--1150, 2003.

\bibitem[Stubbendick and Ibrahim(2006)]{Stubbendick2006}
A.L Stubbendick and J.G Ibrahim.
\newblock {Likelihood-based inference with nonignorable missing responses and
  covariates in models for discrete longitudinal data}.
\newblock \emph{Statistica Sinica}, 16\penalty0 (4):\penalty0 1143--1167, 2006.

\bibitem[Tudor et~al.(2000)Tudor, Koch, and Catellier]{Tudor2000}
Gail~E. Tudor, Gary~G. Koch, and Diane Catellier.
\newblock {Statistical methods for crossover designs in bioenvironmental and
  public health studies}.
\newblock \emph{Handbook of Statistics}, 18:\penalty0 571--614, 2000.

\bibitem[Wei and Tanner(1990)]{Wei1990}
Greg~C.G. Wei and Martin~A. Tanner.
\newblock {A Monte Carlo Implementation of the EM Algorithm and the Poor Man's
  Data Augmentation Algorithm}.
\newblock \emph{Journal of the American Statistical Association}, 85\penalty0
  (411):\penalty0 699--704, 1990.

\end{thebibliography}
\newpage
\section{Appendix}
\subsection{An overview of Monte Carlo EM algorithm}\label{secmcem}
As a method for estimating the parameters, we used the Monte Carlo EM algorithm (MCEM), a variation of the EM algorithm. It is well known that in the presence of missing data, the EM algorithm (\citet{DEMP1977}) is a generic and iterative approach to finding maximum likelihood estimates. There are two key steps in the process: the Expectation step and the Maximization step. For the purpose of understanding the E and M-steps, let the complete data vector be denoted by $\bm{y}=(\bm{y}_o,\bm{y}_m)$, where $\bm{y}_o$ is the vector corresponding to observed data and $\bm{y}_m$ is corresponding to missing data, and $\bm{\theta}$ is the vector of unknown parameters. The E-step calculates the expected value of complete data log-likelihood, $log f(\bm{y},\bm{\theta})$, with respect to the conditional density of unobserved ($\bm{y}_m$) given observed ($\bm{y}_o$) data and a fixed parameter vector ($\bm{\theta}^{'}$). The expectation step is then written as $E_{\bm{y}_m|\bm{y}_o,\bm{\theta}^{'}} (log f(\bm{y},\bm{\theta}))$ or $E_{\bm{y}_m|\bm{y}_o,\bm{\theta}^{'}} (.)$ in short, also called Q-function. In the M-step, the goal is to maximize the expectation step or the Q-function.

Monte Carlo EM (MCEM) algorithm (\citet{Wei1990}, \citet{casella2001}) is a modification of the EM algorithm for intractable E-steps. In this case, E-steps are estimated using Monte Carlo simulations. If we get the Monte Carlo sample $\bm{u}_1, \bm{u}_2, \ldots, \bm{u}_q$ of size $q$ from the conditional distribution $g(\bm{y}_m|\bm{y}_o, \bm{\theta}^{'})$, the expectation is estimated by the Monte Carlo sum
\begin{equation}
    Q_q(\bm{\theta}|\bm{\theta}^{'})=\frac{1}{q}\sum_{t=1}^{q} log f(\bm{y_o},\bm{u}_t,\bm{\theta}). \label{eq estepe}
\end{equation}
By the law of large numbers, the estimator in \cref{eq estepe} converges to the theoretical expectation $E_{\bm{y}_m|\bm{y}_o,\bm{\theta}^{'}} (.)$. Therefore, the MCEM algorithm replaces the E-step by the estimated quantity.

 A number of authors have discussed the convergence properties of MCEM algorithms. \citet{chan1995} demonstrated that, given a suitable starting value, a sequence of parameter values generated by the Monte Carlo EM algorithm will be arbitrarily close to a maximizer of the observed likelihood. The ergodic theory of Markov chains was used by \citet{Fort2003} to demonstrate the almost sure convergence of a Monte Carlo EM algorithm variation.
\subsection{Conditional distribution of missing responses given the observed cases}\label{secapp}
Suppose, $\bm{y}_{i}=(\bm{y}_{mis,i}^{T},\bm{y}_{obs,i}^{T})^{T}$ where $\bm{y}_{mis,i}$ is the $m_{i} \times 1$ vector  of missing responses with corresponding matrices as $\bm{X}_{i_{m_{i}\times j_1}}^{(1)}$ and $\bm{Z}_{i_{m_{i}\times n_i}}^{(1)}$. Similarly, $\bm{y}_{obs,i}$ is the $l_{i} \times 1$ vector  of observed responses with corresponding matrices as $\bm{X}_{i_{l_{i}\times j_1}}^{(2)}$ and $\bm{Z}_{i_{l_{i}\times n_i}}^{(2)}$. The sum of $m_{i}$ and $l_{i}$ is equal to $pmn_{i}$. Accordingly
\begin{equation}\label{eq: app}
\begin{bmatrix}
\bm{y}_{mis,i}\\
\bm{y}_{obs,i}
\end{bmatrix}=\begin{bmatrix}
\bm{X}_{i_{m_{i}\times j_1}}^{(1)}\\
\bm{X}_{i_{l_{i} \times j_1}}^{(2)}
\end{bmatrix}\bm{\beta}+\begin{bmatrix}
\bm{Z}_{i_{m_{i}\times n_i}}^{(1)}\\
\bm{Z}_{i_{l_{i} \times n_i}}^{(2)}
\end{bmatrix}\bm{b_{i}}+\begin{bmatrix}
\bm{e}_{m_{i}\times 1}^{(1)}\\
\bm{e}_{l_{i}\times 1}^{(2)}
\end{bmatrix}; \; \;j_1=p+t+m-2.\end{equation}
The random effects ($\bm{b}_i$) and random errors ($\bm{e}_i$) in \cref{eq: app} follows a normal distribution, so
$$
\bm{y}_{i}= \begin{bmatrix}
\bm{y}_{mis,i}\\
\bm{y}_{obs,i}
\end{bmatrix} \sim N\left(\begin{bmatrix}
\bm{X}_{i}^{(1)} \bm{\beta}\\
\bm{X}_{i}^{(2)} \bm{\beta}
\end{bmatrix}, \begin{bmatrix}
\bm{\Sigma}_{11}&\bm{\Sigma}_{12}\\
\bm{\Sigma}_{21}&\bm{\Sigma}_{22}
\end{bmatrix}\right),$$
where,
$$ \begin{bmatrix}
\bm{\Sigma}_{11}&\bm{\Sigma}_{12}\\
\bm{\Sigma}_{21}&\bm{\Sigma}_{22}
\end{bmatrix}=\begin{bmatrix}
\bm{Z}_{i}^{(1)}\bm{D}_{i}{\bm{Z}_{i}^{(1)}}^T&\bm{Z}_{i}^{(1)} \bm{D}_{i}{\bm{Z}_{i}^{(2)}}^T\\
\bm{Z}_{i}^{(2)} \bm{D}_{i}{\bm{Z}_{i}^{(1)}}^T& \bm{Z}_{i}^{(2)} \bm{D}_{i}{\bm{Z}_{i}^{(2)}}^T
\end{bmatrix}+\sigma_{e}^2 \begin{bmatrix}
\bm{I}_{m_{i}}&0\\
0&\bm{I}_{l_{i}}
\end{bmatrix}.$$ 
Therefore, given the observed cases, the conditional distribution of missing responses is as follows: \begin{equation*}\label{eq:ymis giv y obs}
\bm{y}_{mis,i}|\bm{y}_{obs,i},\bm{\gamma}^{(t)} \sim \pN_{m_{i}}(\bm{X}_{i}^{(1)} \bm{\beta}+\bm{\Sigma}_{12}\bm{\Sigma}_{22}^{-1}(\bm{y}_{obs,i}-\bm{X}_{i}^{(2)} \bm{\beta}),\bm{\Sigma}_{11}-\bm{\Sigma}_{12}\bm{\Sigma}_{22}^{-1}\bm{\Sigma}_{21}).
\end{equation*}
\subsection{Process of generating a sample from the conditional distribution of data considered as missing given the observed data}\label{imputesamp}
We apply Gibbs sampling (\citet{geman1984}, \citet{gelfrad}) to generate a sample from $[\bm{b}_{i},\bm{y}_{mis,i}|\hat{\bm{\gamma}},\bm{y}_{obs,i}]$, i.e., the conditional distribution of unobserved data given observed data. This is a widely used MCMC algorithm that is a special case of the Metropolis-Hastings algorithm. It is used when the joint distribution cannot be directly sampled or when it is difficult to do so, but the conditional distribution of each variable can be calculated. In our case, it is possible to draw samples from the conditional distributions,
  $\bm{b}_i|\hat{\bm{\gamma}},\bm{y}_{mis,i}, \bm{y}_{obs, i}$, and $\bm{y}_{mis, i}|\hat{\bm{\gamma}},\bm{y}_{obs,i}, \bm{b}_i$. Thus, Gibbs sampling is used in this case.
  
The conditional distribution $\bm{b}_i|\bm{y}_{mis,i}, \bm{y}_{obs, i}$ can be obtained from \cref{b giv y} in \Cref{sec mle missing}. 
 In light of \cref{eq: app} of \Cref{secapp}, we note that given $\bm{b}_i$, missing responses and observed responses are independent. Therefore,
$$\bm{y}_{mis,i}|\bm{y}_{obs,i}, \bm{b}_{i}\sim N_{m_i}(\bm{X}_{i}^{(1)} \bm{\beta}+\bm{Z}_{i}^{(1)}\bm{b}_i,\sigma_{e}^2\bm{I}_{m_{i}}).$$
Below are the steps that we follow;  $\hat{\bm{\gamma}}=(\hat{\bm{\beta}},\hat{\sigma_{e}}^2,\hat{\sigma_{s}}^2)$ are the MCEM estimates.
\begin{enumerate}
\item[(i)]  Initial estimates $({\bm{b}_{i}}_{0}, {\bm{y}_{mis,i}}_{0})$ were taken from the respective marginal distributions i.e. $\bm{b}_{i} \sim N(\bm{0},\sigma_{s}^2 \bm{I}_{n_{i}})$ and  \\ $\bm{y}_{mis,i}$ $\sim N(\bm{X}_{i}^{(1)} \bm{\beta},\bm{Z}_{i}^{(1)}\bm{D}_{i}{\bm{Z}_{i}^{(1)}}^T+\sigma_{e}^2 \bm{I}_{m_{i}})$.
\item[(ii)] At the $k^{th}(k \geq 1)$ iteration  $({\bm{b}_{i}}_{k}, {\bm{y}_{mis,i}}_{k})$ were sampled using ${\bm{b}_{i}}_{k}$ from $\bm{b}_{i}|\bm{y}_{obs,i},{\bm{y}_{mis,i}}_{k-1}$ and ${\bm{y}_{mis,i}}_{k}$ from ${\bm{y}_{mis,i}}|\bm{y}_{obs,i},{\bm{b}_{i}}_{k}$. \end{enumerate}
 Step (ii) was repeated 2000 times, and after a burn-in, the average of the remaining samples was used for the final analysis.
\subsection{Additional simulation study results}\label{secsim}
\begin{sidewaystable}[!htp]
\scriptsize
\begin{center}
\caption{\textcolor{black}{Simulation results for 300 data sets with 30 subjects per sequence ($\phi_0=\phi_2=0.1, \phi_1=-0.41$), the overall missing percentage is 38.7: maximum likelihood estimates, SEs, relative bias, and p-values for proposed analysis, complete cases, and imputation analysis. True parameter values are given in parentheses.}}
\label{table simulated datan30}
\begin{tabular}{llllrlllrlllr}
\hline
                  & \multicolumn{4}{c}{Proposed analysis}   & \multicolumn{4}{c}{Complete cases}      & \multicolumn{4}{c}{Imputation analysis}              \\ \hline
Parameter         & Estimate & SE     & Rel\_bias & p-value & Estimate & SE     & Rel\_bias & p-value & Estimate & SE     & Rel\_bias & p-value              \\ \hline
Intercept   (2.5) & 2.5256   & 0.0957 & 0.0102    & $\le$0.0001  & 2.5669   & 0.2181 & 0.0267    & $\le$0.0001 & 2.7468   & 0.1555 & 0.0987    & $\le$0.0001               \\
Period$_1$ (0.4)    & 0.3944   & 0.0933 & -0.0140   & $\le$0.0001 & 0.4324   & 0.2251 & 0.0810    & 0.0552  & 0.2577   & 0.1600 & -0.3557   & 0.1075               \\
Period$_2$   (1.06) & 1.0169   & 0.1142 & -0.0407   & $\le$0.0001 & 1.1060   & 0.2764 & 0.0434    & 0.0001  & 0.8096   & 0.1959 & -0.2362   & $\le$0.0001               \\
Trt$_1$ (0.26)      & 0.2386   & 0.0933 & -0.0823   & 0.0107  & 0.2566   & 0.2251 & -0.0133   & 0.2549  & 0.2218   & 0.1600 & -0.1468   & 0.1659              \\
Trt$_2$ (0.32)      & 0.2986   & 0.1142 & -0.0670   & 0.0091  & 0.3069   & 0.2764 & -0.0409   & 0.2672  & 0.2782   & 0.1959 & -0.1308   & 0.1560               \\
Res$_1$ (0.5)        & 0.4790   & 0.0968 & -0.0421   & $\le$0.0001  & 0.5667   & 0.2363 & 0.1333    & 0.0168  & 0.4203   & 0.1686 & -0.1593   & 0.0128               \\
Res$_2$ (0.7)        & 0.6845   & 0.0968 & -0.0222   & $\le$0.0001  & 0.7023   & 0.2363 & 0.0033    & 0.0031  & 0.5980   & 0.1686 & -0.1457   & 0.0004               \\
Res$_3$ (0.6)        & 0.5830   & 0.0968 & -0.0284   & $\le$0.0001  & 0.6046   & 0.2363 & 0.0076    & 0.0108 & 0.5097   & 0.1686 & -0.1505   & 0.0026               \\
$\sigma_e^2$ (1.44)     & 1.4175   & 0.0589 & -0.0156   & $\le$0.0001  & 1.5937   & 0.2537 & 0.1067    & $\le$0.0001  & 1.3677   & 0.1653 & -0.0502   & $\le$0.0001              \\
$\sigma_s^2$ (0.49)     & 0.4967   & 0.0785 & 0.0136    & $\le$0.0001  & 0.4635   & 0.0821 & -0.0540   & $\le$0.0001  & 0.3823   & 0.1321 & -0.2197   & 0.0039  \\ \hline            
\end{tabular}
\end{center}
\end{sidewaystable}
\end{document}